\documentclass[aps,prb,twocolumn,superscriptaddress,showkeys,citeautoscript]{revtex4-1}

\usepackage[colorlinks=true,bookmarks=false,citecolor=blue,linkcolor=red,hyperfootnotes=true,urlcolor=blue]{hyperref}

\usepackage{physics}
\usepackage{graphicx}
\usepackage{dcolumn}
\usepackage{bm}
\usepackage[dvipsnames]{xcolor}
\usepackage{wasysym}
\usepackage{enumitem}
\usepackage{booktabs} 
\usepackage{gensymb}
\usepackage{xcolor}
\usepackage{amssymb}
\usepackage{booktabs}
\usepackage{array}
\newcolumntype{L}{>{$}l<{$}}
\newcolumntype{C}{>{$}c<{$}}
\newcolumntype{R}{>{$}r<{$}}

\def\vecx{\mathbf{x}}

\makeatletter
\def\p@subsection{}
\makeatother

\newcommand{\vecb}{\mathbf{b}}

\newcommand{\vecq}{\mathbf{q}}
\newcommand{\vecn}{\mathbf{n}}

\newcommand{\veck}{\mathbf{k}}

\newcommand{\code}[1]{\texttt{#1}}
\newcommand{\ii}{\code{i}}
\newcommand{\tb}[1]{\tilde{#1}}

\usepackage{hyperref}

\DeclareRobustCommand{\emailaddr}[1]{%
  \href{mailto:#1}{\nolinkurl{#1}}%
}

\begin{document}

\title{Finite-temperature Green's function cluster expansion from thermofield doubles:\\Breakdown of the polaron picture}

\author{Matthew R. Carbone}
\thanks{These authors contributed equally.}
\affiliation{Computational Science Initiative, Brookhaven National Laboratory, Upton, New York 11973, USA}

\author{Stepan Fomichev}
\thanks{These authors contributed equally.}
\affiliation{Department of Physics and Astronomy, University of British Columbia, Vancouver, British Columbia V6T 1Z1, Canada}
\affiliation{Stewart Blusson Quantum Matter Institute, University of British Columbia, Vancouver, British Columbia V6T 1Z4, Canada}

\author{Benedikt Kloss}
\affiliation{Center for Computational Quantum Physics, Flatiron Institute, 162 5th Avenue, New York, New York 10010, USA}

\author{Andrew J. Millis}
\affiliation{Department of Physics, Columbia University, New York, New York 10027, USA}
\affiliation{Center for Computational Quantum Physics, Flatiron Institute, 162 5th Avenue, New York, New York 10010, USA}

\author{Mona Berciu}
\affiliation{Department of Physics and Astronomy, University of British Columbia, Vancouver, British Columbia V6T 1Z1, Canada}
\affiliation{Stewart Blusson Quantum Matter Institute, University of British Columbia, Vancouver, British Columbia V6T 1Z4, Canada}

\author{David R. Reichman}
\affiliation{Department of Chemistry, Columbia University, New York, New York 10027, USA}

\author{John Sous}\thanks{Author to whom correspondence should be addressed: \emailaddr{john.sous@yale.edu}}
\affiliation{Department of Applied Physics, Yale University, New Haven, Connecticut 06511, USA}
\affiliation{Energy Sciences Institute, Yale University, West Haven, Connecticut 06516, USA}

\date{\today}

\begin{abstract}
We introduce a method, numerically exact in principle, for computing the momentum- and frequency-resolved single-particle Green's function of a polaron at finite temperature. The method, which we refer to as the finite-temperature Green's function cluster expansion, combines two ingredients: the generalized Green's function cluster expansion, a numerically exact extension of the momentum average family of methods that solves the polaron problem at zero temperature through a hierarchy of equations of motion for restricted phonon cloud configurations; and the thermofield double formalism, which maps the thermal trace onto a pure-state expectation value over a doubled Hilbert space. The resulting equations of motion have the same algebraic structure as those of the multi-boson zero-temperature theory, with the temperature entering through a Bogoliubov-type mixing angle that controls the coupling to a set of fictitious bath bosons. We implement the method in our open-source software package and benchmark it on the one-dimensional Holstein polaron, computing spectral functions, dispersions, lifetimes, and effective masses across coupling regimes and temperatures up to $T/\Omega \sim 1$. Where finite-temperature density matrix renormalization group results are available, we find quantitative agreement at affordable computational cost. The method recovers momentum-resolved spectra directly in frequency space, with no time evolution or analytic continuation. We also discuss the practical costs of the approach. In particular, since the doubled phonon Hilbert space has a non-trivial configuration structure in which real and fictitious clouds compete, convergence in the corresponding cloud parameters requires care.
\end{abstract}

\maketitle

\section{Introduction}
\label{sec:intro}

The accurate treatment of strong electron-phonon coupling presents a serious challenge to a full \emph{ab initio} description of materials. In contrast to the electronic structure problem, for which density functional theory (DFT) has enabled large-scale simulations with reasonable accuracy, the treatment of electron-phonon coupling has mostly proceeded through lower-complexity effective models, with the hope that an embedding scheme akin to DFT plus dynamical mean-field theory (DFT+DMFT) can ultimately link such models to first-principles electronic Hamiltonians.

Here we focus on an important limiting class of electron-phonon problems: the polaron problem, in which a single carrier --- an electron or hole injected into an otherwise empty band --- is dressed by its coupling to the lattice into an emergent quasiparticle, the polaron. Beyond its role as the canonical testing ground for electron-phonon methods, the polaron limit directly describes momentum-resolved spectroscopy and transport in lightly doped insulators and semiconductors, where the carrier density is low enough that carrier-carrier interactions can be neglected.

A broad family of numerical methods has been developed to study the polaron problem in effective models such as those of the Holstein~\cite{holstein1959studiesI,holstein1959studiesII} and Peierls/Su-Schrieffer-Heeger~\cite{su1979solitons,su1980soliton} type, ranging from exact diagonalization (ED) and variational ED~\cite{bonvca1999holstein,FehskeVED}, to continuous-time Monte Carlo~\cite{kornilovitch1998continuous}, diagrammatic Monte Carlo (DMC)~\cite{mishchenko2000diagrammatic,prokof1998polaron}, density matrix renormalization group (DMRG) methods~\cite{Jeckelmann2,jeckelmann1998density,BK}, and the variational momentum-average (MA) approximation and its generalizations~\cite{berciu2006green,goodvin2006green,berciu2007systematic,berciu2010momentum,carbone2021numerically,carbone2022generalized}. Most of these approaches were originally formulated at zero temperature. New methods are required to access the physically relevant regime in which phonons retain their quantum character but the lattice has non-vanishing thermal occupation, i.e.\ when the ratio of temperature to a characteristic phonon frequency satisfies $T/\Omega \lesssim 1$. This is the regime that controls charge transport in molecular and organic semiconductors at and near room temperature, polaron mobilities in halide perovskites and other soft-lattice materials, and the breakdown of simple quasiparticle descriptions in correlated systems with intermediate-frequency lattice modes.

Existing approaches to the Holstein model at finite temperature include ED of small systems~\cite{de1997dynamical}, continued-fraction-based methods~\cite{paganelli2006tunnelling}, dynamical mean-field theory~\cite{ciuchi1997dynamical,mitric2022spectral}, DMC~\cite{mishchenko2015mobility}, finite-$T$ Lanczos~\cite{bonvca2019spectral}, and finite-temperature DMRG~\cite{jansen2020finite,jansen2022finite}. Each of these comes with trade-offs. ED is limited by Hilbert space dimension at any non-trivial coupling. Finite-$T$ Lanczos can suffer from finite-size effects. DMC is formulated in imaginary time and requires analytic continuation to access real-frequency spectra. DMFT in its single-site form does not provide direct access to momentum-resolved quantities, and is in any case best suited to systems at non-zero carrier density rather than to the single-carrier limit of interest here; cluster extensions ameliorate the momentum-resolution issue at significant additional cost. Finite-$T$ DMRG, in our view the most mature of these methods, is computationally intensive and proceeds in real time, with $A(k,\omega)$ obtained from a Fourier transform of finite-time correlation functions.

In this paper, we propose a complementary method that returns $A(k,\omega)$ directly in frequency space at any chosen $k$ in the Brillouin zone, without time evolution or analytic continuation. We do so by extending the generalized Green's function cluster expansion (GGCE) of Ref.~\cite{carbone2021numerically} to finite temperature and combining it with the thermofield double (TFD) formalism~\cite{takahashi1996thermo,umezawa1982thermo}. The resulting method, which we refer to as the finite-temperature Green's function cluster expansion (TGCE), inherits from GGCE the central physical idea of the momentum-average family: that the dominant phonon configurations dressing a charge carrier form spatially compact clouds, and that an expansion organized by cloud size converges rapidly even in the strong-coupling regime where on-site phonon occupations may become large. In TGCE the role of the cloud is taken over by a doubled cloud whose configuration space spans both physical and auxiliary (``fictitious'') bath modes; temperature enters as a Bogoliubov mixing angle that couples the two species.

The remainder of the paper is organized as follows. Section~\ref{sec:formalism} reviews the GGCE method at $T=0$, summarizes the TFD construction, and combines them to derive the TGCE equations of motion. Section~\ref{sec:implementation} describes the implementation in the open-source GGCE Python package~\cite{carbone2022generalized}, with attention to the parts of the implementation that are specific to the doubled Hilbert space. Section~\ref{sec:performance} characterizes the matrix scaling and sparsity, compares the three available linear-algebra backends, and presents convergence studies in the cloud parameters. Section~\ref{sec:results} demonstrates the method on the one-dimensional Holstein model, computing temperature- and momentum-resolved spectral functions, polaron dispersions, lifetimes, and effective masses, and comparing to finite-$T$ DMRG where data is available. Section~\ref{sec:conclusion} summarizes the results and discusses the limitations and outlook.

\section{Formalism}
\label{sec:formalism}

\subsection{Review of GGCE}
\label{sec:formalism:GGCE}

Consider a mobile charged particle, such as an electron or hole, coupled to dispersive phonons. The most general single-carrier Hamiltonian of interest here takes the form
\begin{multline}
\label{eq:H_general}
    H = H_0 + V = \sum_\veck \varepsilon_\veck c_\veck^\dagger c_\veck + \hbar \sum_\vecq \Omega_\vecq b_\vecq^\dagger b_\vecq \\
    + \sum_{\veck \vecq} g(\veck, \vecq)\, c_{\veck + \vecq}^\dagger c_\veck \left( b_{-\vecq}^\dagger + b_\vecq \right),
\end{multline}
where $\varepsilon_\veck$ is the carrier dispersion and $\Omega_\vecq$ the phonon dispersion, with $c_\veck^\dagger$ and $b_\vecq^\dagger$ the corresponding creation operators; $g(\veck, \vecq)$ is the electron-phonon vertex, which in the most general case depends on both the carrier momentum $\veck$ and the phonon momentum $\vecq$, as for instance in the Peierls model. The non-interacting part $H_0$ contains the kinetic and phononic terms; the interaction $V$ renders $H$ non-integrable.

\begin{figure}[b]
\centering
\includegraphics[width=0.8\linewidth]{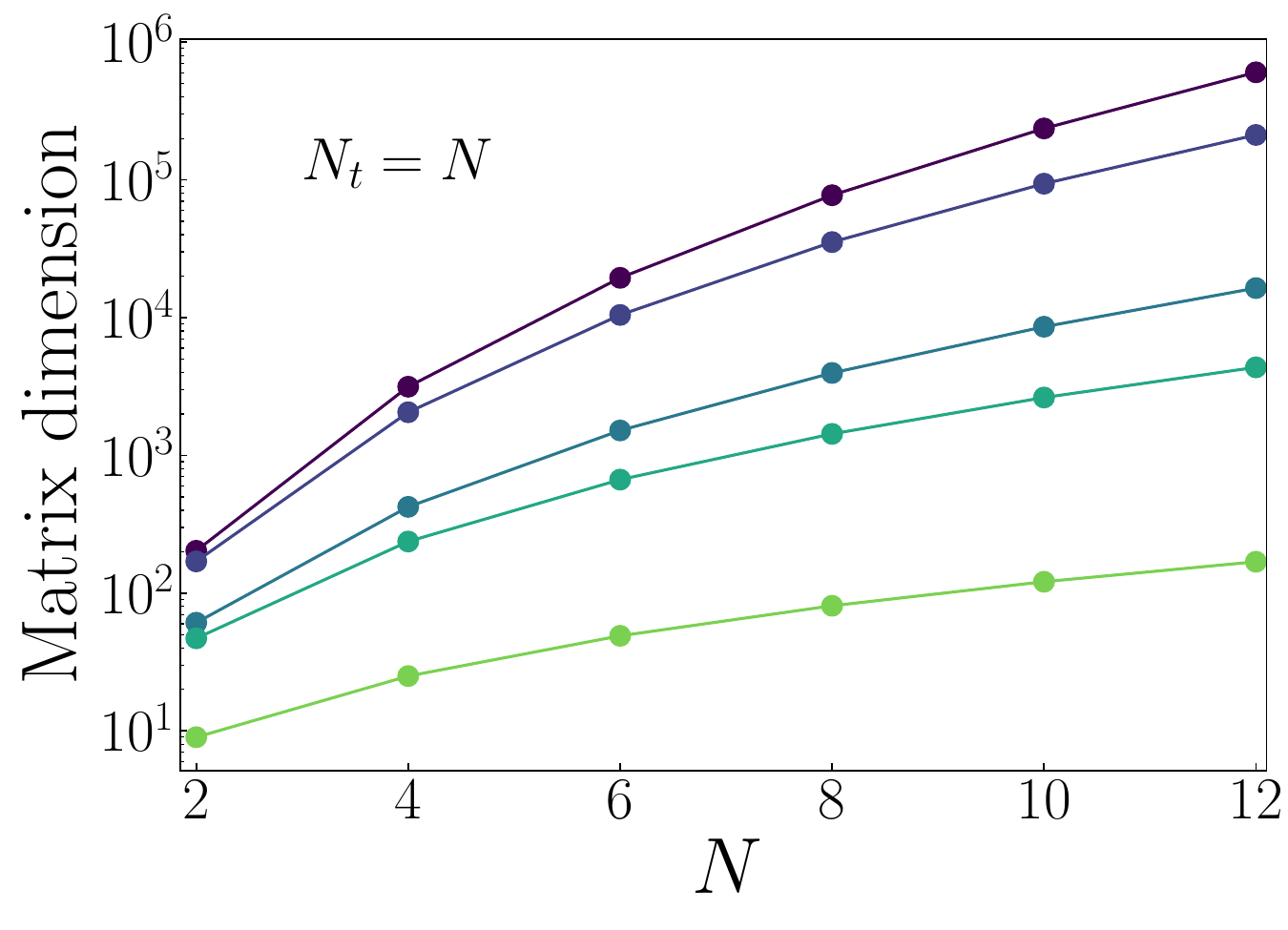}\\
\includegraphics[width=0.8\linewidth]{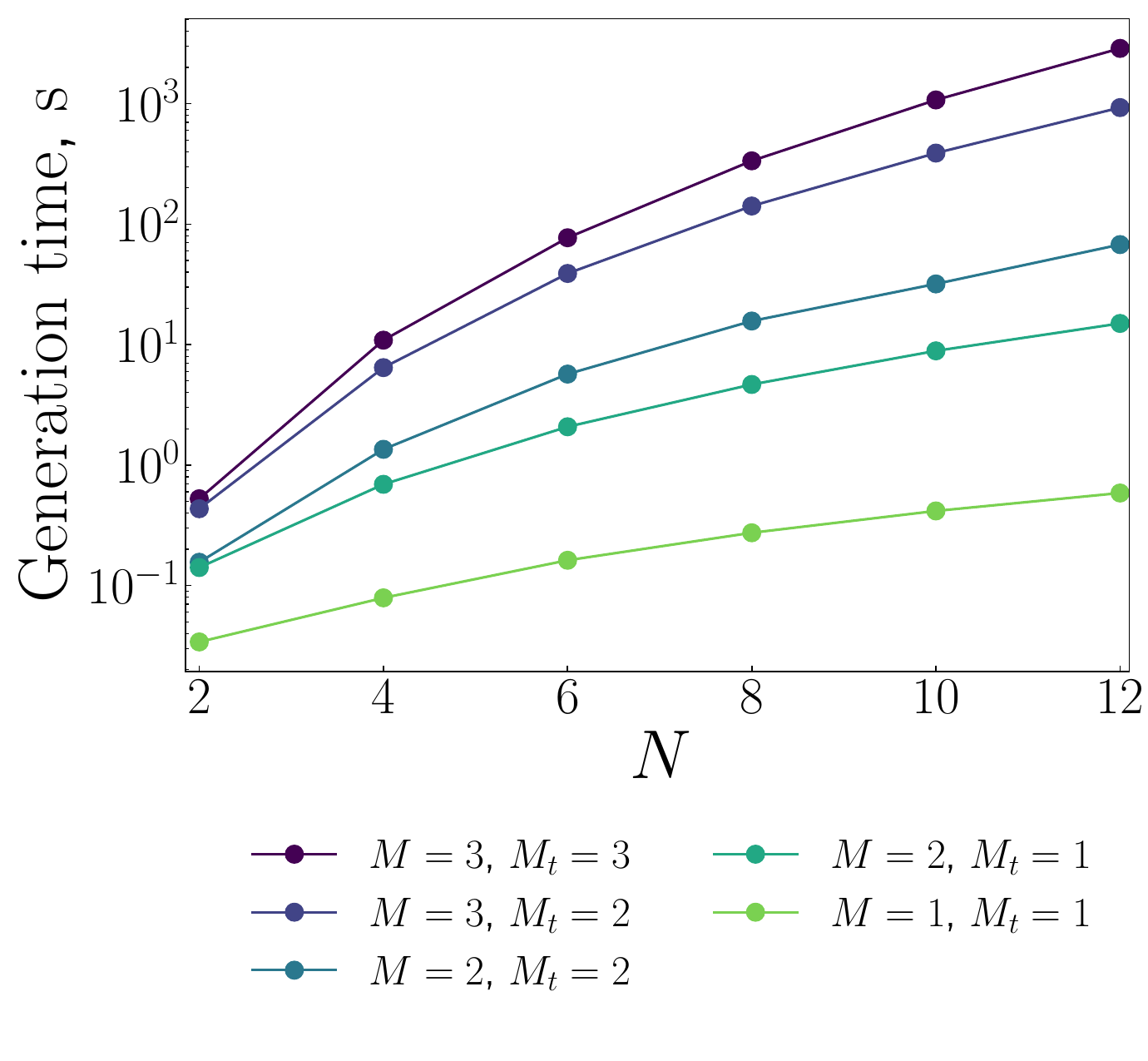}
\caption{(a) Size of the matrix containing the full equations of motion for the Holstein polaron, as a function of the cloud parameters. We fix $N_t = N$ and show representative curves for several values of the real ($M$) and fictitious ($M_t$) cloud extents. (b) Equation-generation time as a function of $N$. Both quantities exhibit power-law scaling with $N$, as in zero-temperature GGCE, with the doubling of the phonon species entering as a prefactor.}
\label{fig:matrscaling_size}
\end{figure}

\begin{figure*}
\centering

\begin{minipage}[t]{0.45\linewidth}
  \centering
  \vspace{0pt}
  \includegraphics[
    width=\linewidth,
    trim=0 0 0 0,
    clip
  ]{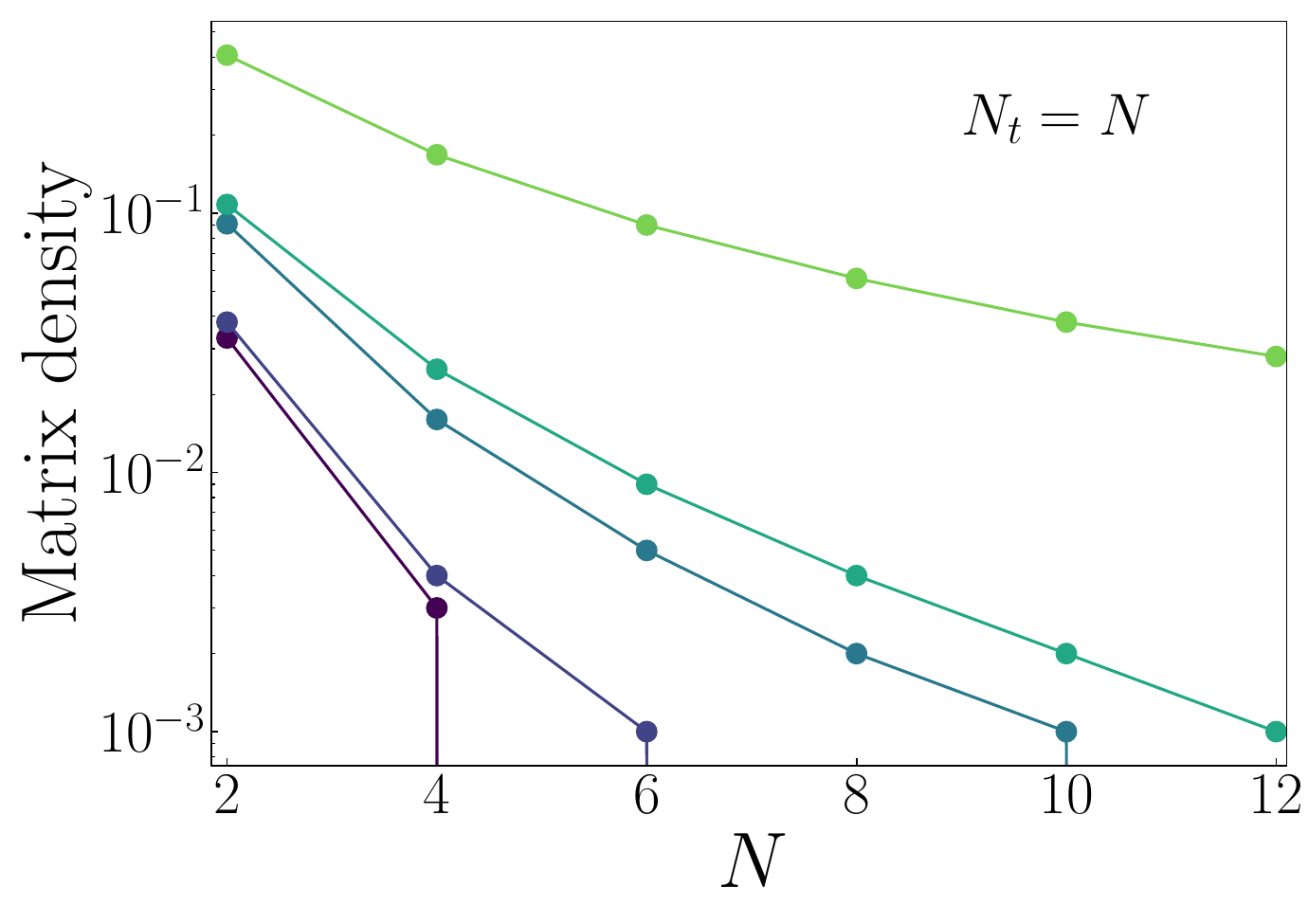}
\end{minipage}
\hspace{0.7cm}
\begin{minipage}[t]{0.45\linewidth}
  \centering
  \vspace{0pt}
  \includegraphics[
    width=\linewidth,
    trim=1mm 0 0 0,
    clip
  ]{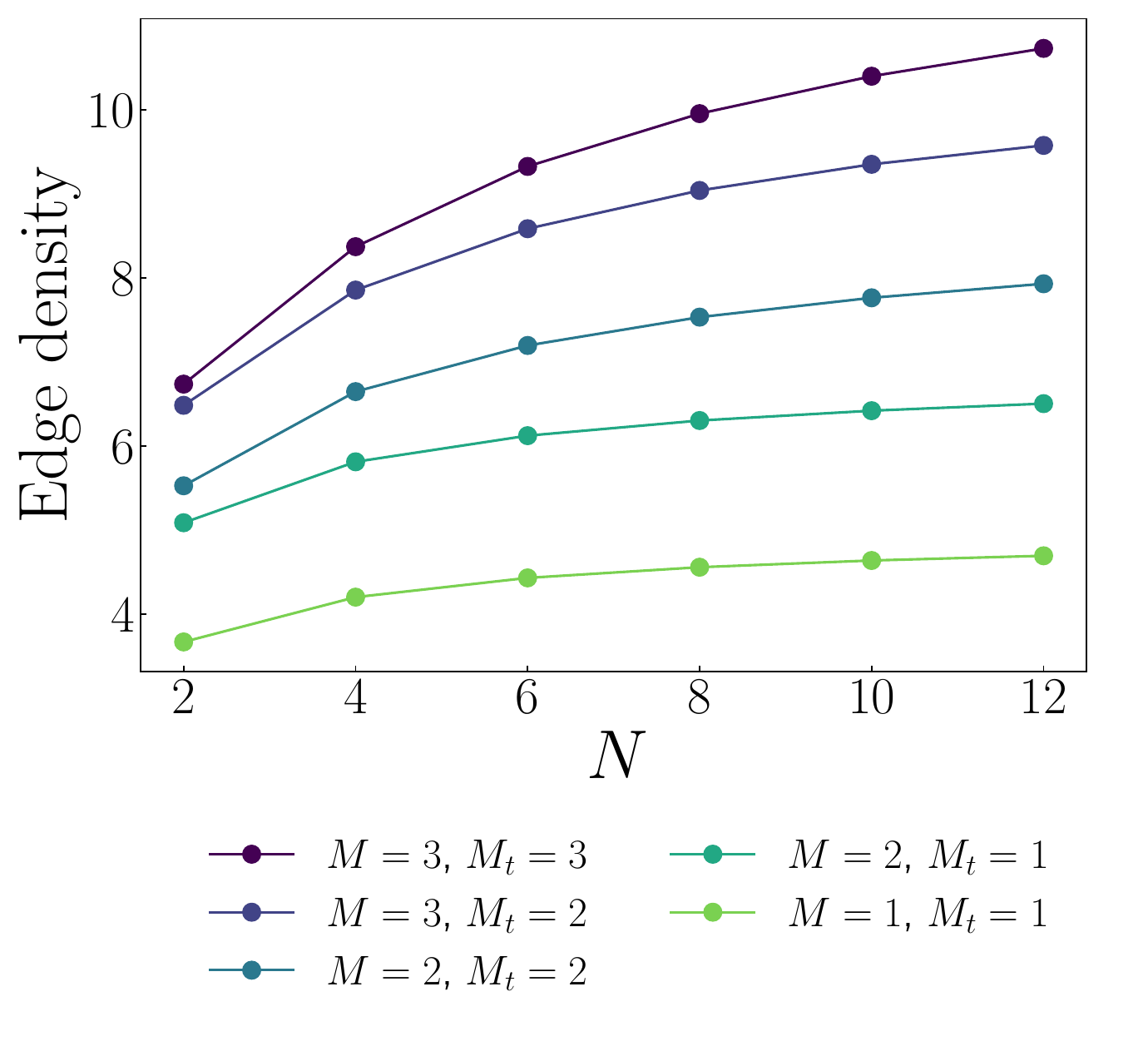}
\end{minipage}

\vspace{-4mm}


\begin{minipage}[t]{0.45\linewidth}
  \centering
  \vspace{0pt}
  \includegraphics[width=\linewidth]{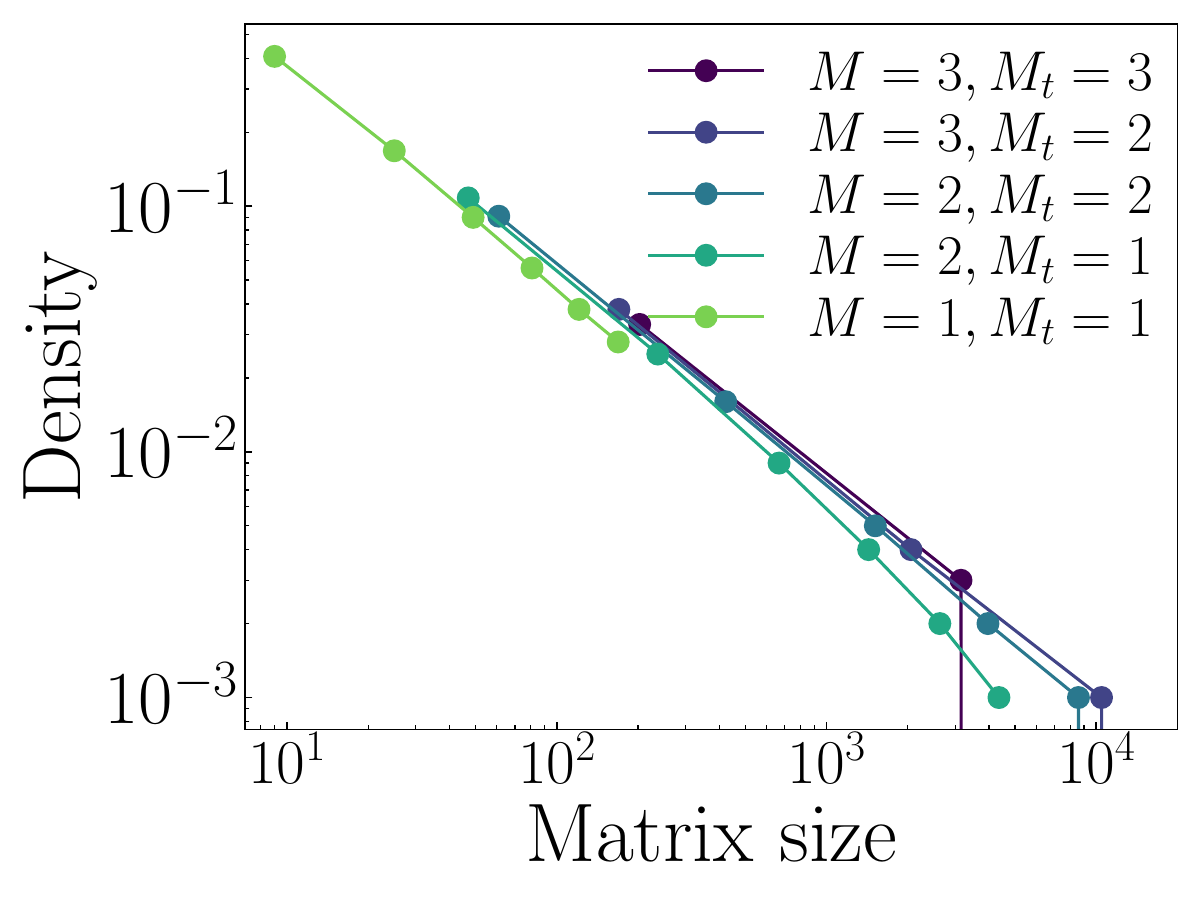}
\end{minipage}
\hspace{0.7cm}
\begin{minipage}[t]{0.48\linewidth}
  \centering
  \vspace{20pt}
  \includegraphics[width=\linewidth]{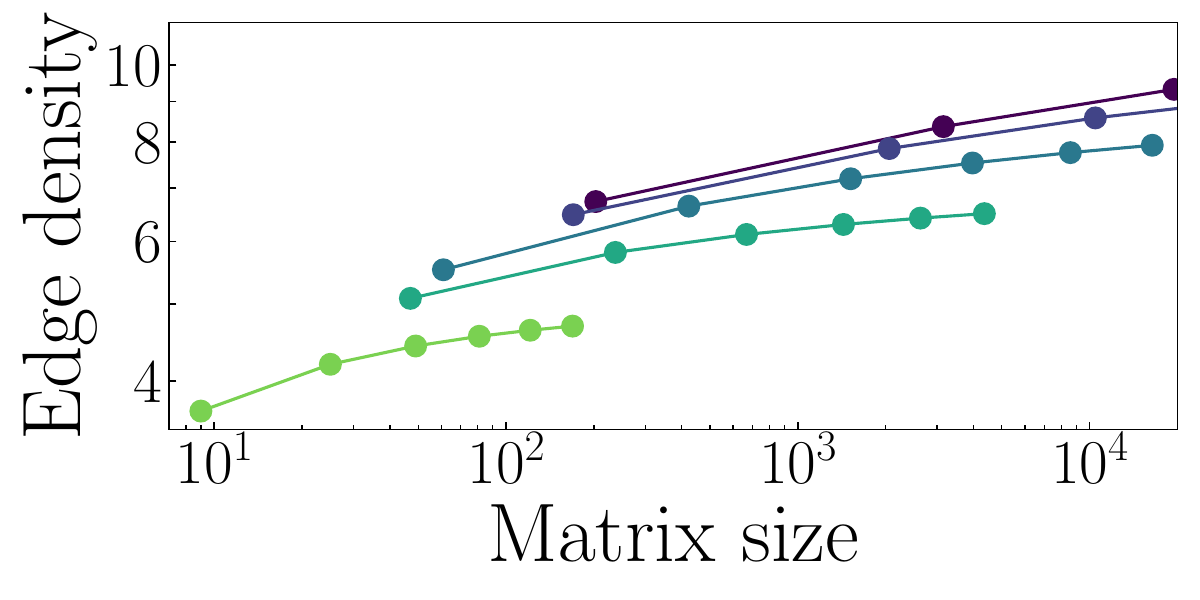}
\end{minipage}
\caption{(a) Matrix density, defined as the number of non-zero entries divided by the total number of matrix entries, as a function of the cloud cutoff $N$. (b) Edge density, $d_e \equiv \mathrm{nnz}/\sqrt{\mathrm{size}}$, as a function of $N$. (c, d) The same quantities, plotted as functions of the matrix size. The density falls rapidly with system size, but the edge density grows, indicating that the number of non-zero entries scales between linearly and quadratically in the matrix dimension. See Sec.~\ref{sec:performance:scaling} for the role of these quantities in characterizing the linear-solve cost.}
\label{fig:matrscaling_sparsity}
\end{figure*}

\begin{figure*}
\centering
\includegraphics[width=0.325\linewidth]{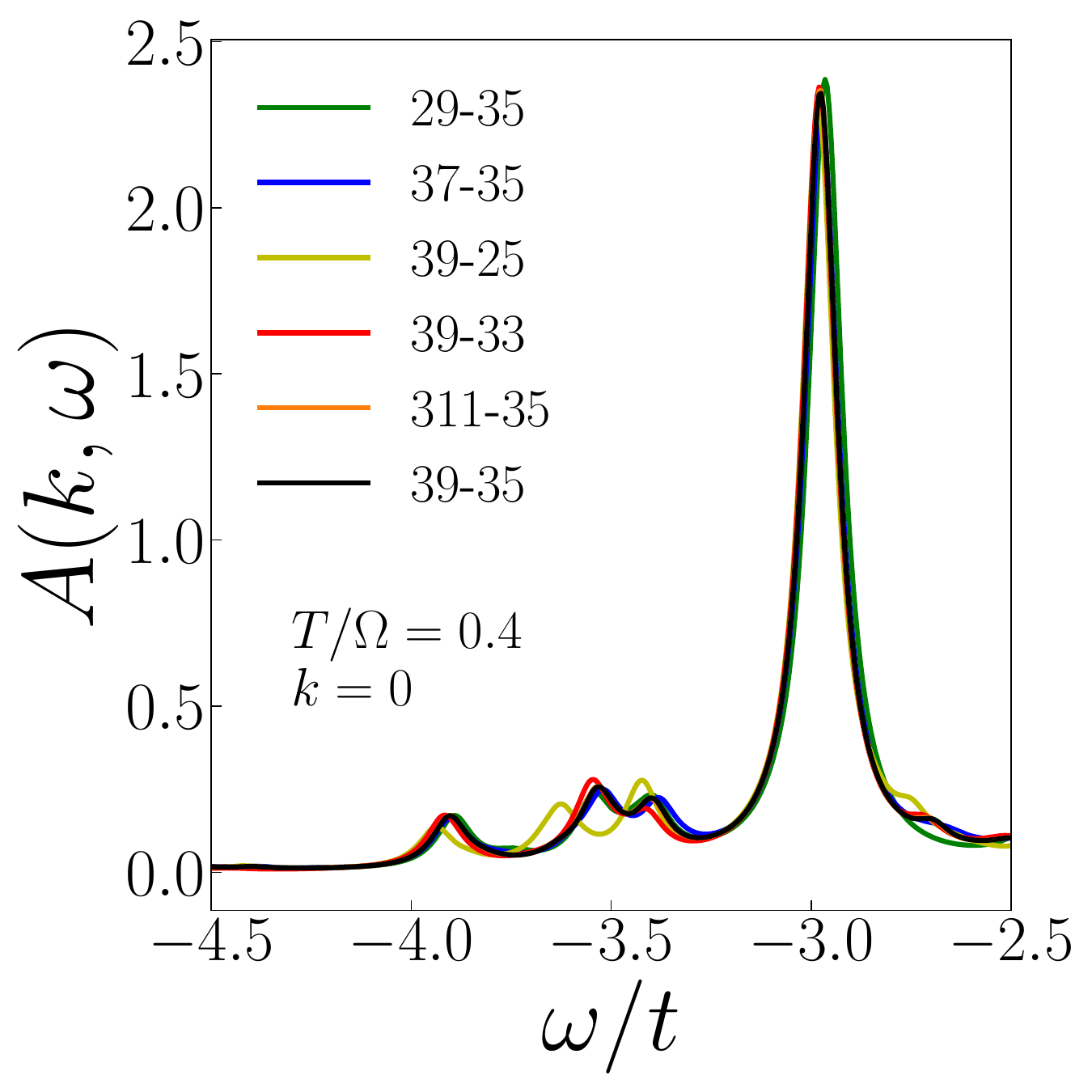}\hspace{1.2cm}
\includegraphics[width=0.325\linewidth]{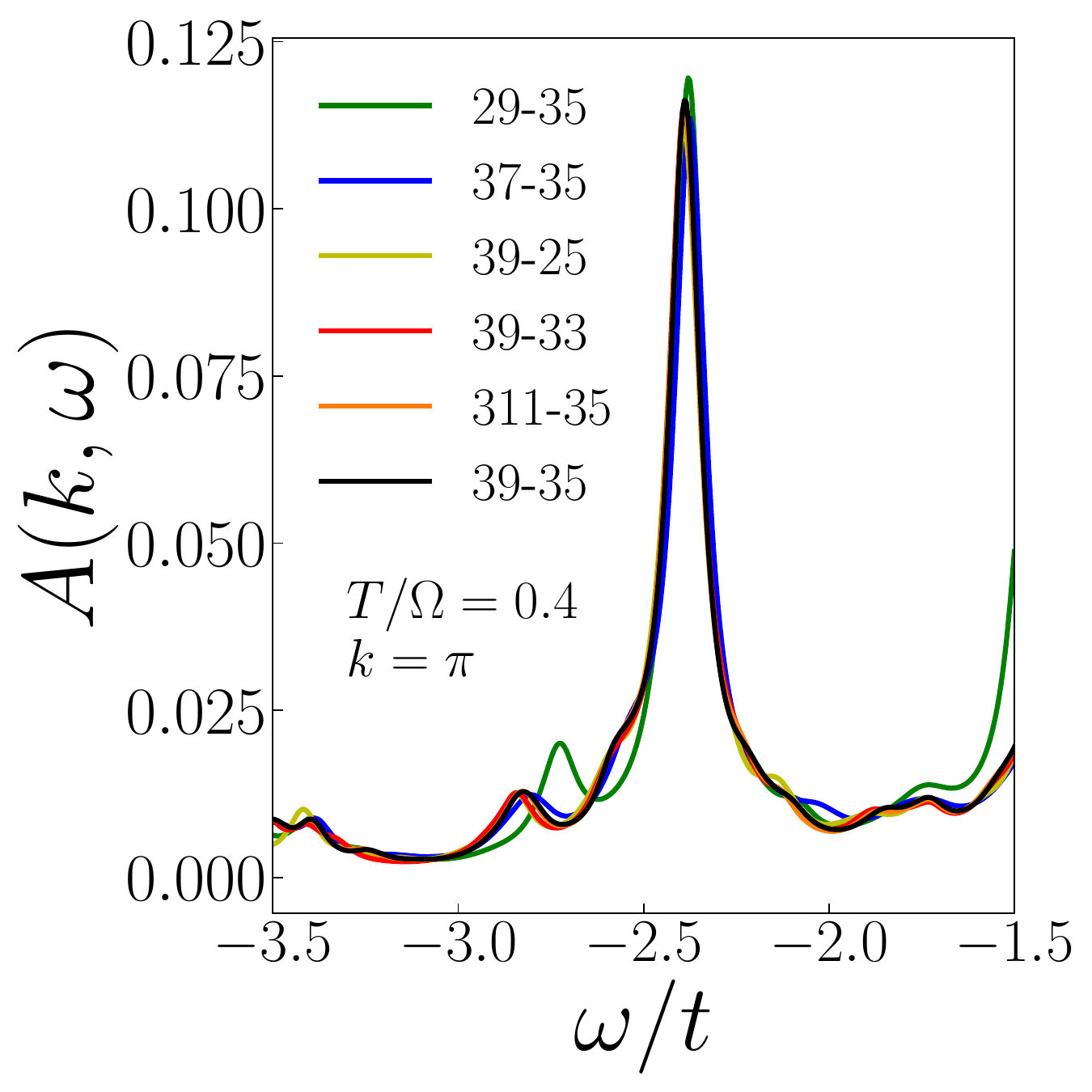} \\
\vspace{5mm}
\includegraphics[width=0.325\linewidth]{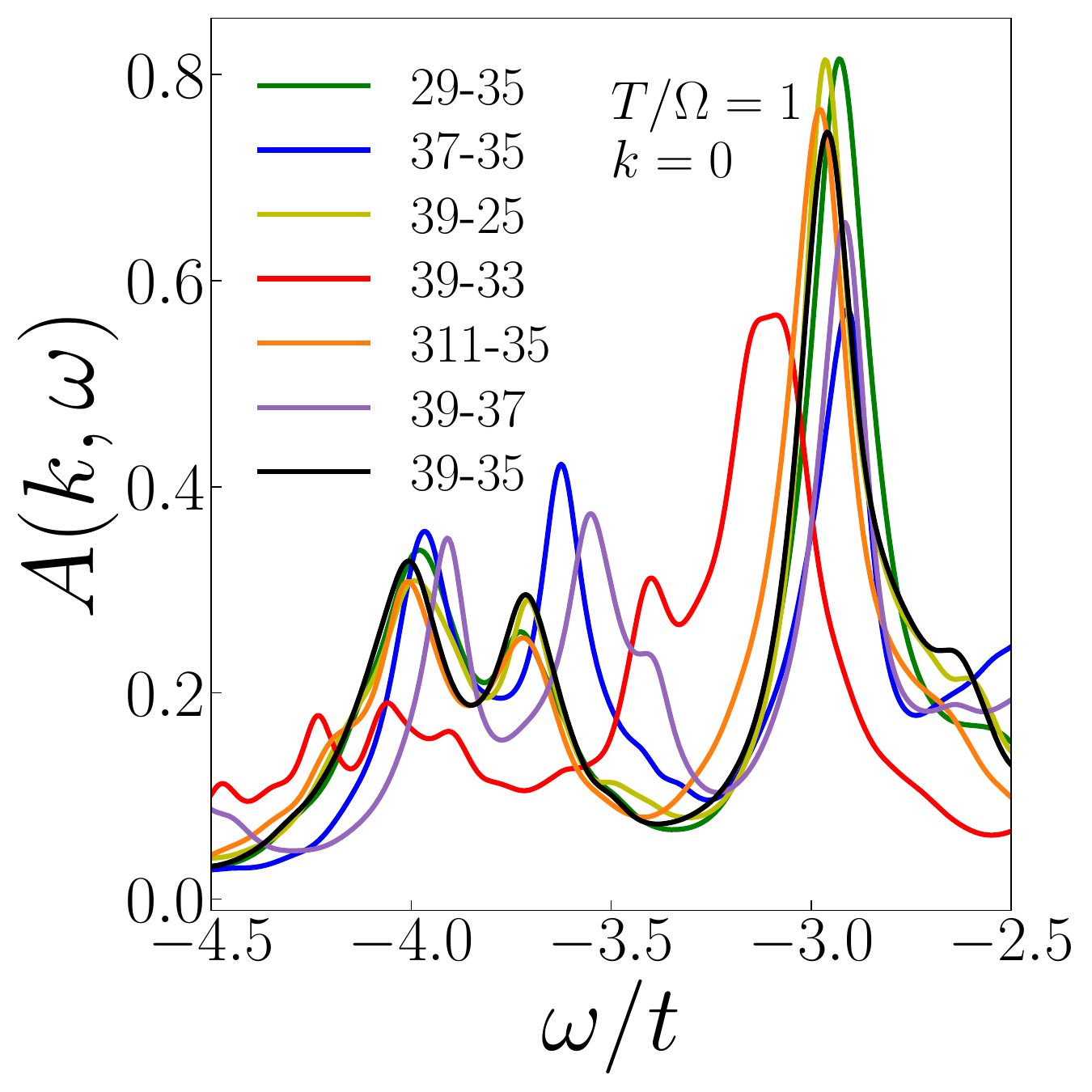}\hspace{1.2cm}
\includegraphics[width=0.325\linewidth]{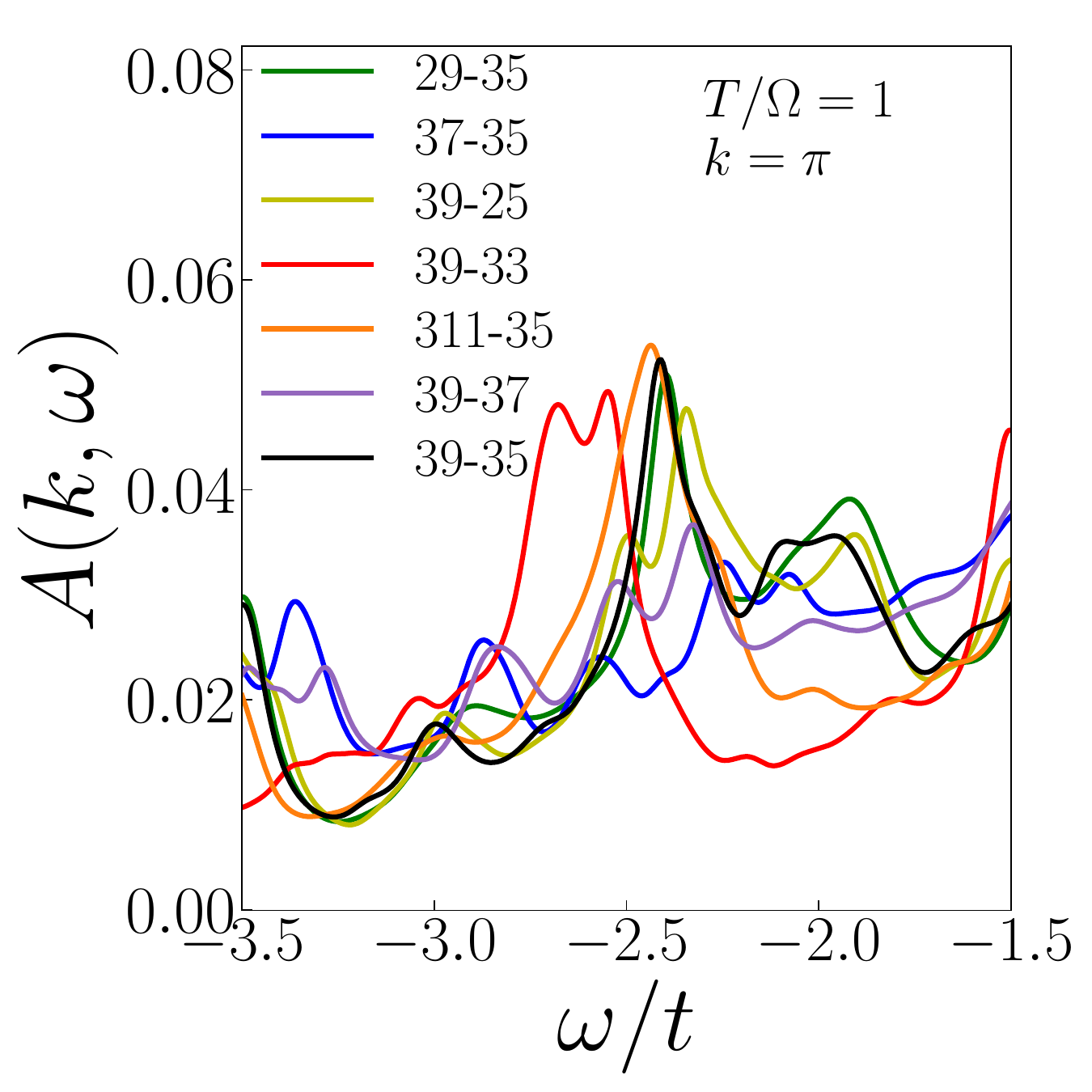}
\caption{Convergence of the spectral function in the cloud parameters $(M, N, M_t, N_t)$ at $\lambda = 1$, $\Omega = 1$, for $T/\Omega = 0.4$ (top row) and $T/\Omega = 1.0$ (bottom row), at $k = 0$ (left column) and $k = \pi$ (right column). Curves are labeled $M N$-$M_t N_t$. The polaron peak is well-converged at intermediate temperature for $M N$-$M_t N_t = 39$-$35$ and adjacent configurations; at $T/\Omega = 1$ convergence is slower but the peak position and width remain stable.}
\label{fig:convergence}
\end{figure*}

The GGCE approach~\cite{carbone2021numerically,carbone2022generalized}, a numerically exact generalization of the MA family of methods~\cite{berciu2006green,berciu2007systematic,berciu2010momentum}, computes the fully-interacting single-particle Green's function
\begin{equation}
\label{eq:G_def}
    G(\veck, \omega) = \mel{0}{c_\veck\, G(\omega)\, c_\veck^\dagger}{0},
\end{equation}
where the propagator
\begin{equation}
\label{eq:G_propagator}
    G(\omega) = [\omega - H + \ii \eta]^{-1}
\end{equation}
is evaluated at real frequency $\omega$ with infinitesimal broadening $\eta$. Repeated application of the Dyson identity $G = G_0 + G\, V\, G_0$, with $G_0(\omega) = (\omega - H_0 + \ii \eta)^{-1}$ the non-interacting propagator, generates a hierarchical set of equations of motion for the generalized propagators
\begin{equation}
\label{eq:f_def}
    f_\vecn(\veck, \omega; \delta) = \frac{1}{\sqrt{\mathcal{N}}} \sum_i e^{\ii \veck\cdot \mathbf{R}_i}\, \bra{0} c_\veck\, G(\omega)\, c_{i-\delta}^\dagger\, B_{i,\vecn}^\dagger \ket{0},
\end{equation}
where the cloud creation operator $B^\dagger_{i,\vecn}$ is defined explicitly as
\begin{equation}
\label{eq:B_explicit}
    B^\dagger_{i,\vecn} = \prod_{j} \frac{(b^\dagger_{i+j})^{n_j}}{\sqrt{n_j!}},
\end{equation}
and creates a normalized cloud of phonons of profile $\vecn$ anchored at site $i$, with $n_j$ phonons on the site at relative position $j$ from the anchor. The equations relating $f_\vecn$ to $f_{\vecn^\pm}$ (clouds with one phonon added or removed) are exact; truncating them at some total cloud size $N = \sum_j n_j$ on a spatial window of $M$ sites recovers exact diagonalization in the limit $N, M \to \infty$.

The non-trivial feature of GGCE and its MA ancestor is that the truncation is organized in a physically motivated way. Rather than imposing a constant phonon-number cutoff per site, as is necessary in ED- or DMRG-style approaches, GGCE caps both the \emph{spatial extent} $M$ of the cloud and the \emph{total phonon number} $N$ within it. For a single carrier dressed by a local cloud of phonons, this matches the physical structure of the polaron problem: in the strong-coupling regime the number of phonons per site can be large, but the cloud itself remains relatively compact. As a result, GGCE generally converges at modest $(M, N)$ even in regimes where ED is intractable, and has been benchmarked against alternative numerically exact methods across coupling and phonon-frequency regimes for both Holstein and Peierls models~\cite{carbone2021numerically,carbone2021bond}.

The GGCE as formulated above, however, is intrinsically a zero-temperature method: Eq.~\eqref{eq:G_def} is an expectation value in the carrier-empty, phonon-empty vacuum. Extending it to finite temperature requires a route to the thermal Green's function that is structurally compatible with the cloud-truncation machinery just described. This is provided by the thermofield double.

\subsection{The thermofield double}
\label{sec:formalism:TFD}

The thermal expectation value of a time-dependent operator $A(\tau)$ at inverse temperature $\beta$ is given by the trace
\begin{equation}
\label{eq:thermal_trace}
    \expval{A(\tau)}_\beta = \tr\!\left[ A(\tau)\, \rho(\beta) \right], \qquad A(\tau) = e^{\ii H \tau} A(0) e^{-\ii H \tau},
\end{equation}
with $\rho(\beta) = Z(\beta)^{-1} e^{-\beta H}$ and partition function $Z(\beta) = \tr e^{-\beta H}$. Direct evaluation of the trace requires summing over a complete basis of the full carrier-phonon Hilbert space, which is prohibitively expensive once the lattice and phonon spectrum are non-trivial.

The TFD formalism rewrites this trace as a pure-state expectation value in an enlarged (doubled) Hilbert space~\cite{takahashi1996thermo,umezawa1982thermo}. To each state $\ket{n}$ of the original system one associates a fictitious partner $\ket{\tilde n}$ living in an auxiliary copy $\tilde{\mathcal{H}}$ of the Hilbert space; the tilded states commute with all physical operators and serve only as a bookkeeping device. The thermal vacuum
\begin{equation}
\label{eq:TFD_vacuum}
    \ket{\bar 0(\beta)} = \frac{1}{\sqrt{Z(\beta)}} \sum_n e^{-\beta E_n/2}\, \ket{n} \otimes \ket{\tilde n}
\end{equation}
reproduces the thermal trace as $\expval{A(t)}_\beta = \mel{\bar 0(\beta)}{A(t)}{\bar 0(\beta)}$. For polaron problems, the initial density matrix factorizes into a carrier vacuum and a non-interacting phononic Boltzmann weight, since one considers a single carrier injected into a phonon bath. The formulas below are written for dispersionless (Einstein) phonons of frequency $\Omega$, the case relevant to the Holstein model studied in this work; the generalization to dispersive phonons proceeds mode by mode, with $\theta(\beta) \to \theta_\vecq(\beta) = \mathrm{arctanh}\, e^{-\beta \Omega_\vecq/2}$. Under these assumptions, $\ket{\bar 0(\beta)}$ admits an explicit closed-form representation as a unitary (squeezing) transformation acting on the doubled vacuum,
\begin{equation}
\label{eq:TFD_unitary}
    \ket{\bar 0(\beta)} = e^{-\ii G(\beta)}\, \ket{0} \otimes \ket{\tilde 0},
\end{equation}
with
\begin{equation}
\label{eq:TFD_G}
    G(\beta) = \ii\, \theta(\beta) \sum_i \left( b_i^\dagger \tb b_i^\dagger - b_i \tb b_i \right),
\quad
    \theta(\beta) = \mathrm{arctanh}\, e^{-\beta \Omega/2}.
\end{equation}
The squeezing angle $\theta(\beta)$ vanishes as $T \to 0$ ($\beta \to \infty$), and grows monotonically with temperature, recovering the zero-temperature problem in the appropriate limit. A complete derivation of Eqs.~\eqref{eq:TFD_unitary}--\eqref{eq:TFD_G}, including the proof that the squeezing transformation closes within the SU(1,1) Lie algebra spanned by $b^\dagger \tb b^\dagger$, $b \tb b$, and their commutator, is given in Ref~\cite{carbone2021dynamical}.

\subsection{Derivation of TGCE}
\label{sec:formalism:TGCE}

Combining the TFD representation with the equations of motion of Sec.~\ref{sec:formalism:GGCE} requires expressing the thermal Green's function in the doubled Hilbert space in a form that retains the structure of Eq.~\eqref{eq:G_propagator}. The thermal Green's function reads
\begin{equation}
\label{eq:G_thermal_TFD}
    \ii \mathcal{G}(k, \tau; \beta) = \mel{\bar 0(\beta)}{T\, c_k(\tau)\, c_k^\dagger}{\bar 0(\beta)},
\end{equation}
where $T$ is the time-ordering operator and $c_k(\tau) = e^{\ii H \tau}\, c_k\, e^{-\ii H \tau}$. Two manipulations bring this expression to a form structurally compatible with the GGCE machinery. First, since the fictitious vibrational Hamiltonian $\tilde H_{\rm vib} = \Omega \sum_i \tb b_i^\dagger \tb b_i$ commutes with the carrier operators and annihilates the doubled vacuum, an identity insertion $e^{-\ii \tilde H_{\rm vib} \tau}\, e^{+\ii \tilde H_{\rm vib} \tau}$ can be absorbed into the time evolutions, effectively replacing $H \to \bar H \equiv H - \tilde H_{\rm vib}$ in the propagators without changing the value of the matrix element. Second, since the squeezing generator $G(\beta)$ commutes with the carrier operators, inserting $e^{-\ii G(\beta)}\, e^{+\ii G(\beta)} = \openone$ between adjacent operators in the Heisenberg string rotates $\bar H$ into the temperature-dependent thermal Hamiltonian $\bar H(\beta) \equiv e^{\ii G(\beta)}\, \bar H\, e^{-\ii G(\beta)}$. After these manipulations and Fourier transformation, the Green's function takes the compact form
\begin{equation}
\label{eq:G_TFD_final}
    \mathcal{G}(k, \omega; \beta) = \mel{\bar 0}{c_k\, \frac{1}{\omega - \bar H(\beta) + \ii \eta}\, c_k^\dagger}{\bar 0},
\end{equation}
with $\ket{\bar 0} \equiv \ket{0} \otimes \ket{\tilde 0}$ and
\begin{equation}
\label{eq:Hbar}
    \bar H(\beta) = e^{\ii G(\beta)}\, \bar H\, e^{-\ii G(\beta)},
\quad
    \bar H \equiv H - \tilde H_{\rm vib}.
\end{equation}
The subtraction of $\tilde H_{\rm vib}$ from $H$ before the squeezing rotation ensures that the non-interacting part of $\bar H(\beta)$ is diagonal in the doubled phonon basis, with no direct couplings between real and fictitious modes. Eq.~\eqref{eq:Hbar_holstein} below makes this explicit in the Holstein case. It is at this step that the structural compatibility with GGCE becomes visible: $\bar H(\beta)$ has the same operator structure as the original $H$, but with the phononic sector replaced by two non-interacting boson species (the real and fictitious modes) entering with opposite energies, and the electron-phonon vertex split between the two species in a temperature-dependent way.

\begin{figure}[b]
\raggedright
\hspace{5mm}\includegraphics[width=0.75\linewidth]{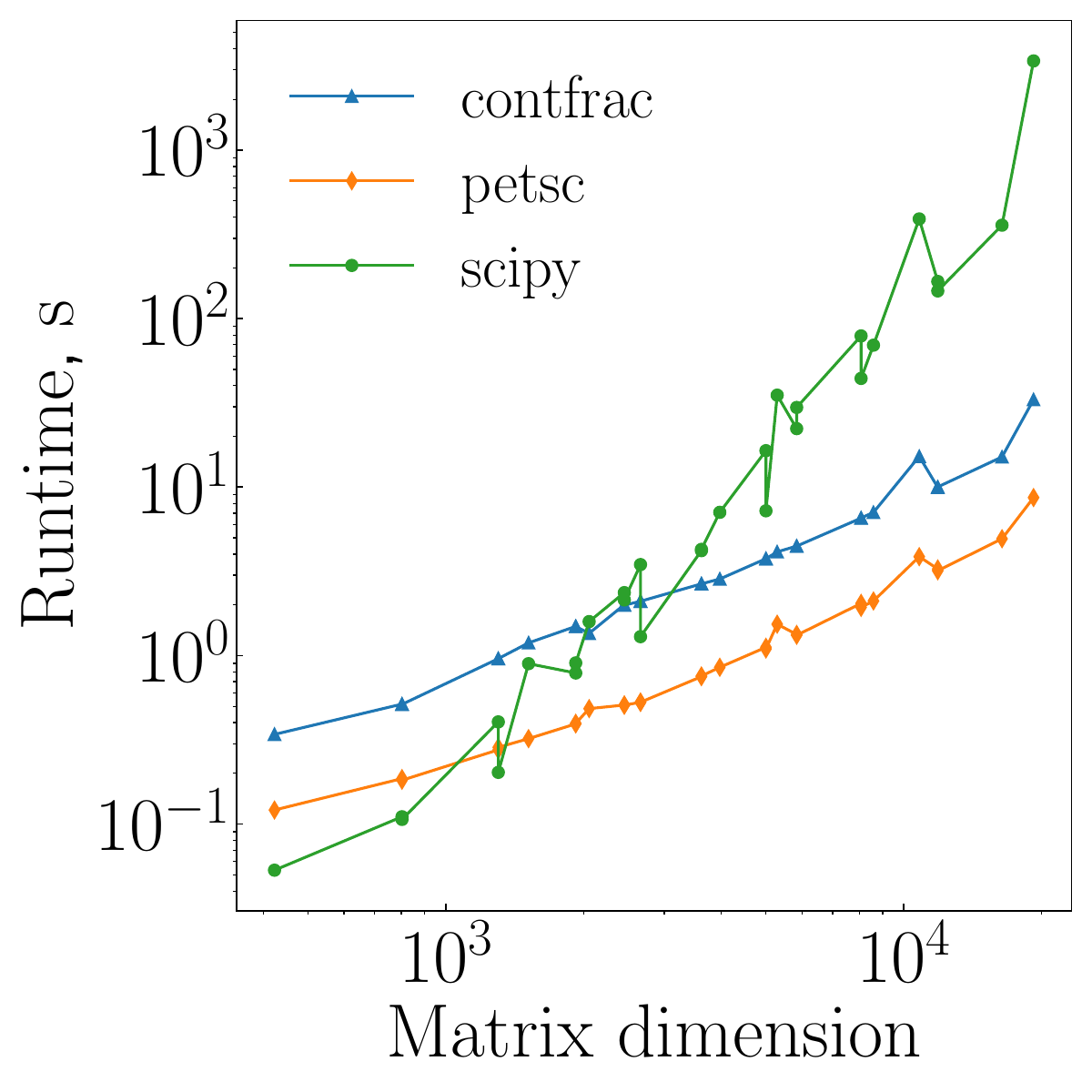}
\caption{Wall-clock time per $(k, \omega)$ point for the three available linear-solver backends in the GGCE package: a SciPy direct solver, a continued-fraction (contfrac) solver, and a PETSc-based sparse solver. Calculations are run on one compute node, distributed across 4 MPI ranks with 8 CPUs per rank. PETSc is fastest for matrix sizes above $\sim 10^3$, where the linear system is no longer canonically sparse.}
\label{fig:solver_runtime}
\end{figure}

Concretely, for the Holstein model with dispersionless phonons of frequency $\Omega$ and on-site coupling $g$ (the momentum-independent limit of the vertex $g(k,q)$ in Eq.~\eqref{eq:H_general}),
\begin{equation}
\label{eq:H_holstein}
    H = K + \Omega \sum_i b_i^\dagger b_i + g \sum_i c_i^\dagger c_i (b_i^\dagger + b_i),
\end{equation}
with $K = -t \sum_{\langle ij \rangle} c_i^\dagger c_j$, the unitary transformation acts on the bosonic operators as
\begin{equation}
\label{eq:bog_transform}
    e^{\ii G(\beta)}\, b_i\, e^{-\ii G(\beta)} = \cosh\theta(\beta)\, b_i + \sinh\theta(\beta)\, \tb b_i^\dagger,
\end{equation}
and analogously for $\tb b_i$. The transformed Hamiltonian becomes
\begin{equation}
\label{eq:Hbar_holstein}
    \bar H(\beta) = K + \Omega \sum_i \!\left( b_i^\dagger b_i - \tb b_i^\dagger \tb b_i \right) + V(\beta),
\end{equation}
with the squeezed interaction
\begin{equation}
\label{eq:V_squeezed}
    V(\beta) = \cosh\theta(\beta)\, V + \sinh\theta(\beta)\, \tilde V,
\end{equation}
where
\begin{equation}
\label{eq:V_holstein}
    V = g \sum_i c_i^\dagger c_i (b_i^\dagger + b_i)
\end{equation}
is the Holstein interaction from Eq.~\eqref{eq:H_holstein} expressed on the real phonon species, and
\begin{equation}
\label{eq:Vt}
    \tilde V = g \sum_i c_i^\dagger c_i (\tb b_i^\dagger + \tb b_i)
\end{equation}
is the same interaction acting on the fictitious species.
Two features deserve emphasis. First, the structure of Eqs.~\eqref{eq:Hbar_holstein}--\eqref{eq:Vt} is identical to that of a two-mode Holstein model with two species of bosons coupling locally to the carrier, except that the vibrational Hamiltonian of the fictitious species enters with reversed sign. Second, the temperature enters only through the two coefficients $\cosh\theta(\beta)$ and $\sinh\theta(\beta)$, which set the relative weight of the carrier's coupling to the real and fictitious sectors. At $T = 0$, $\theta \to 0$, $\sinh\theta \to 0$, the fictitious sector decouples, and one recovers the original zero-temperature GGCE problem; at finite $T$, the fictitious modes acquire weight and dress the carrier alongside the physical phonons.

The equations of motion now follow the same construction as in Sec.~\ref{sec:formalism:GGCE}, with the cloud configuration vector $\vecn$ promoted to a two-row matrix $(n, \tb n)$ keeping track of phonon occupation in both species, and the cloud-extent and cloud-number cutoffs generalized to four parameters, $M$ and $N$ on the real cloud, and $M_t$ and $N_t$ on the fictitious cloud. The free propagator $G_0(\omega)$ acquires a temperature-dependent energy shift: for a state with $n$ real and $\tb n$ fictitious phonons, it returns $G_0(k, \omega - \Omega(n - \tb n))$. The opposite sign of $\tb n$ in the energy denominator is a direct consequence of $\bar H_{\rm vib} = H_{\rm vib} - \tilde H_{\rm vib}$ in Eq.~\eqref{eq:Hbar_holstein}, and as we will discuss in Sec.~\ref{sec:results}, it has practical consequences for the convergence of the method that have no analog in zero-temperature GGCE.

\begin{figure*}[t]
\centering
\includegraphics[width=0.3\linewidth]{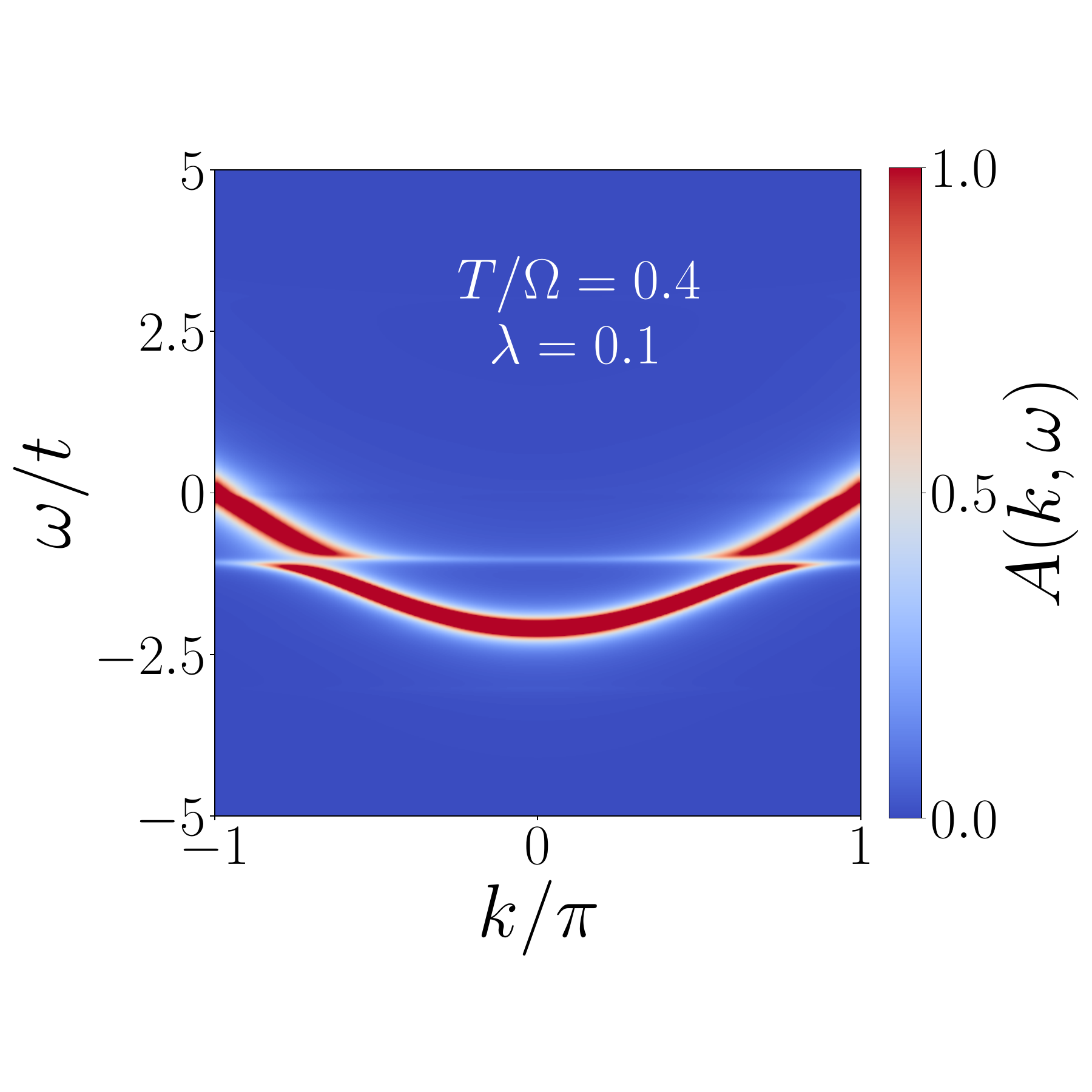}%
\includegraphics[width=0.3\linewidth]{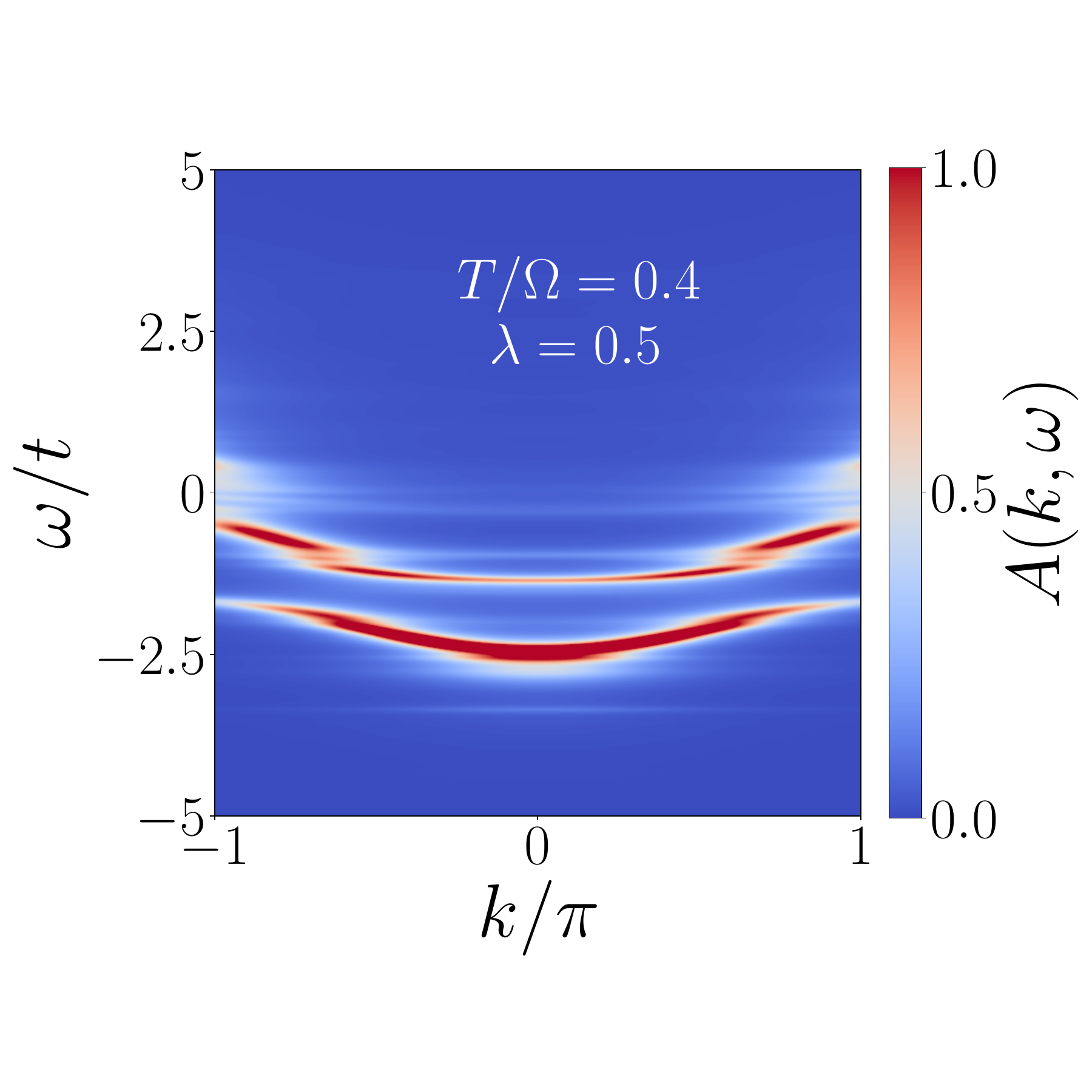}%
\includegraphics[width=0.3\linewidth]{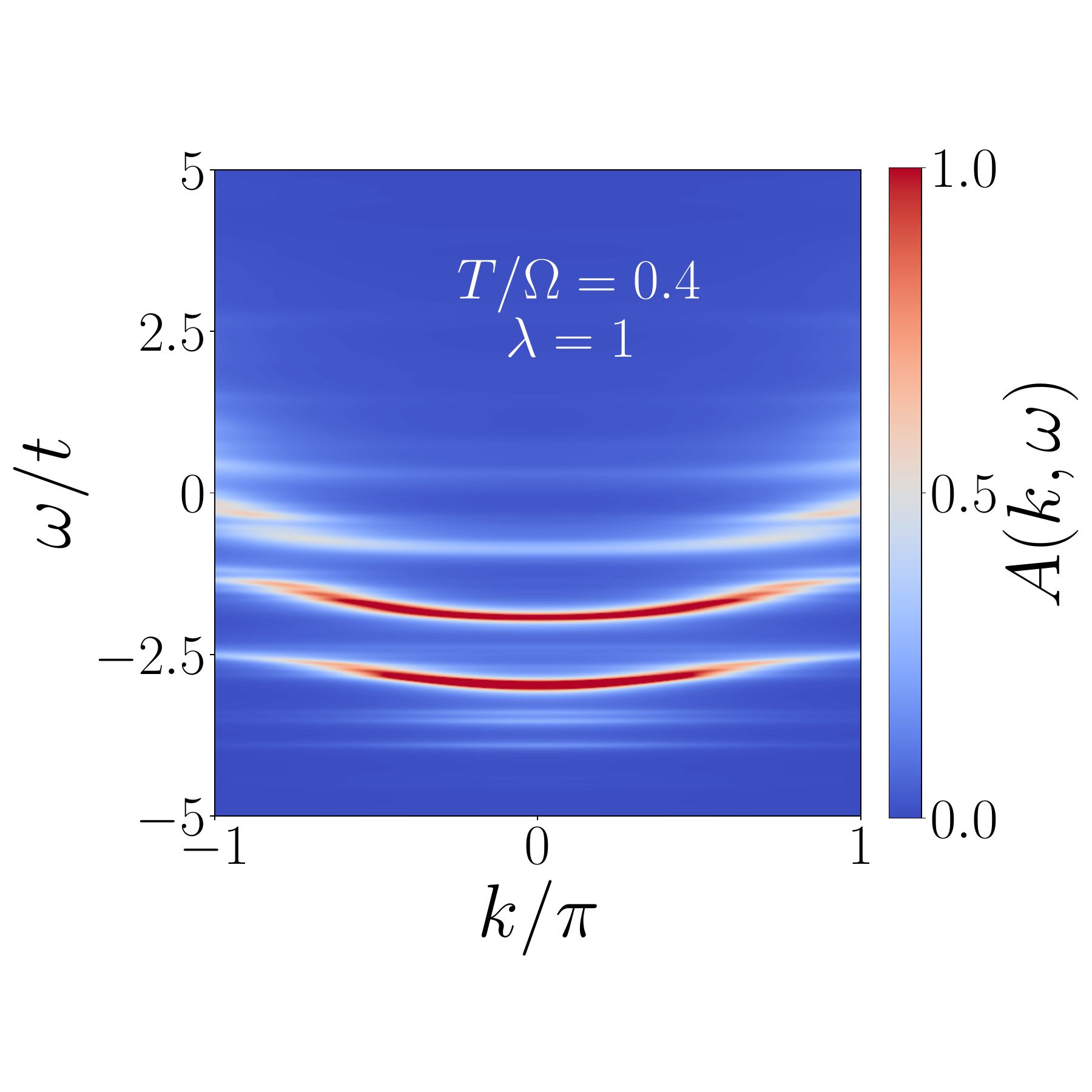}\\
\vspace{-4mm}
\makebox[0.3\linewidth]{\small $\lambda = 0.1$}%
\makebox[0.3\linewidth]{\small $\lambda = 0.5$}%
\makebox[0.3\linewidth]{\small $\lambda = 1$}\\[0.5mm]
\includegraphics[width=0.3\linewidth]{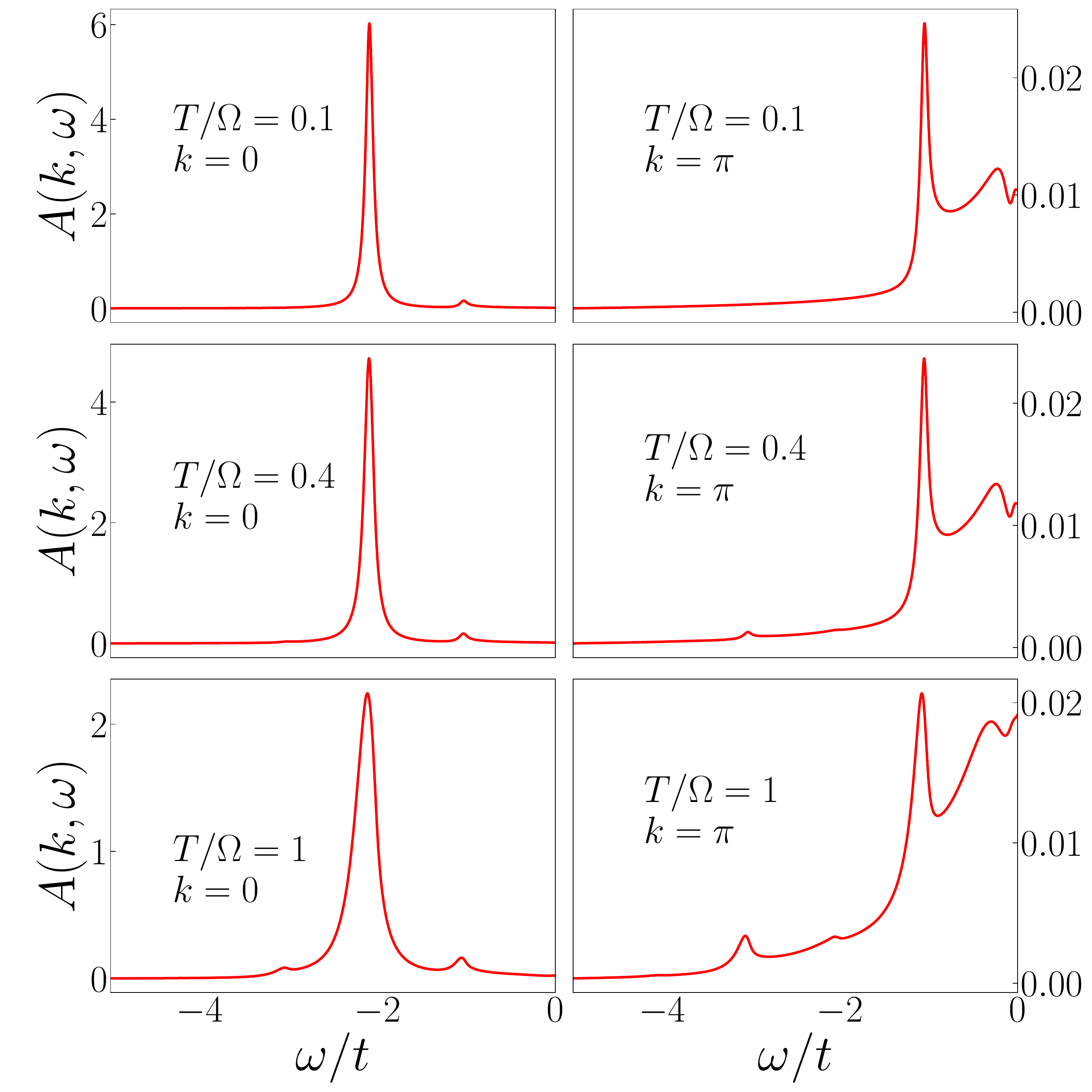}
\includegraphics[width=0.3\linewidth]{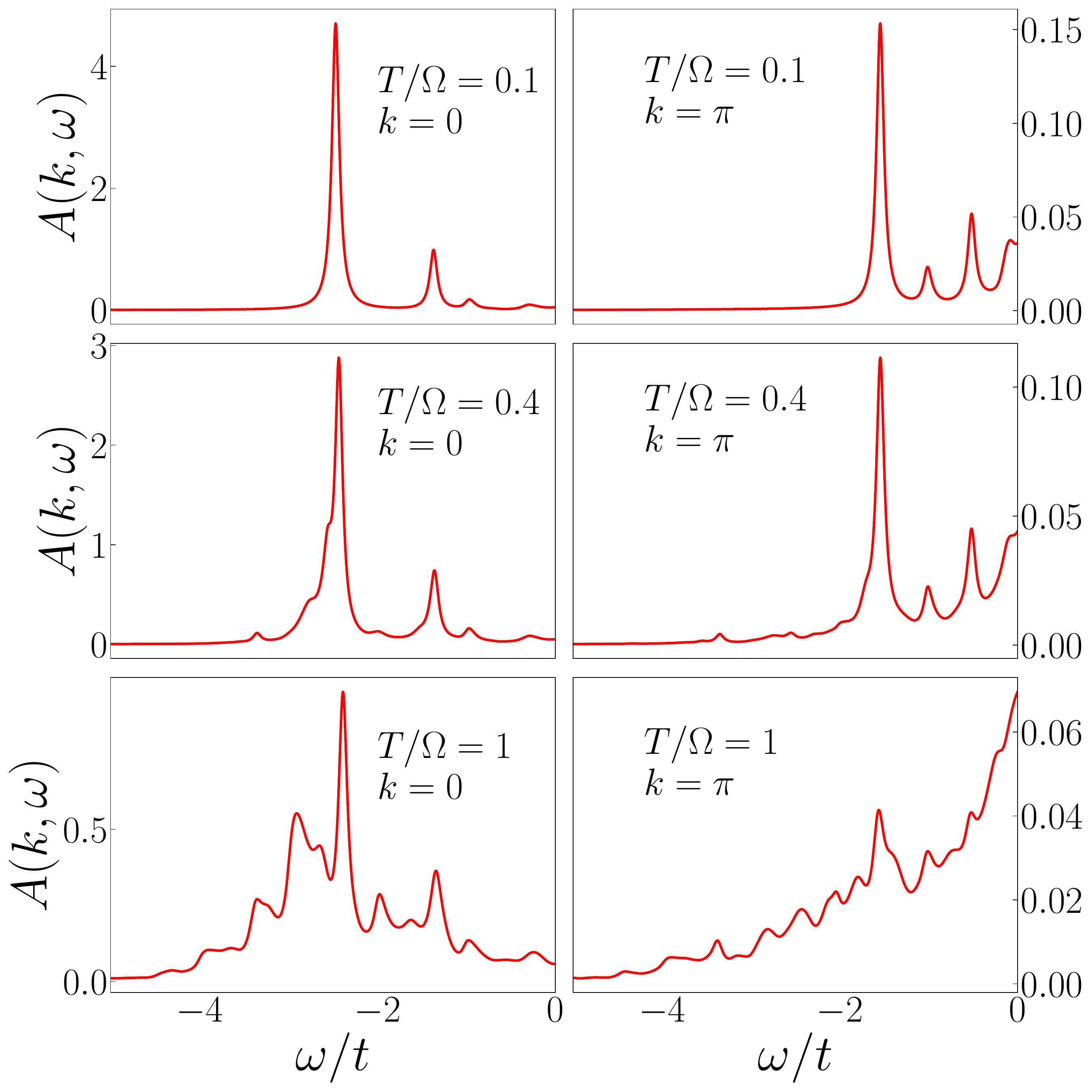}%
\includegraphics[width=0.3\linewidth]{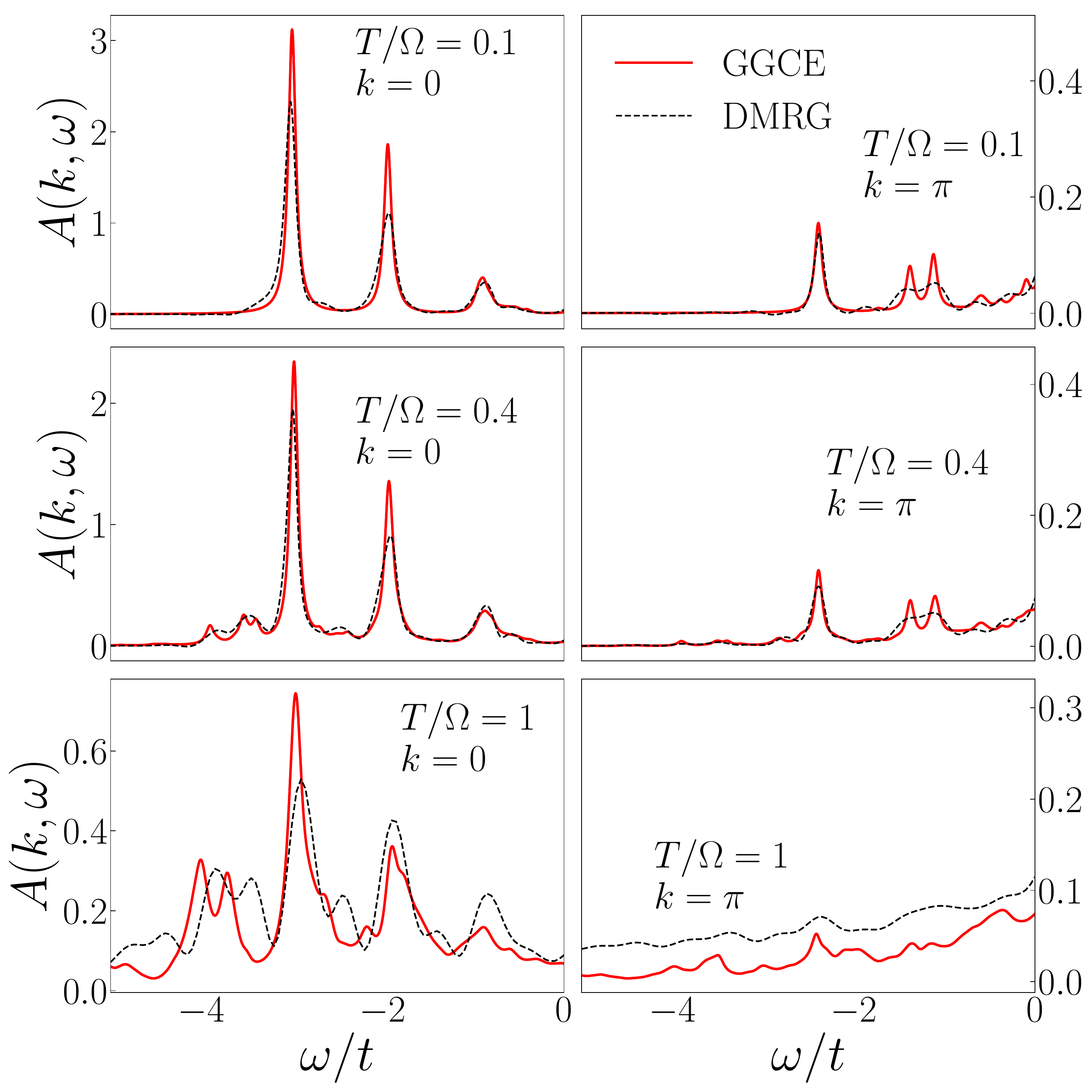}
\caption{(Top row) Spectral functions $A(k, \omega)$ for the one-dimensional Holstein model computed with TGCE at $T/\Omega = 0.4$ and varying coupling: $\lambda = 0.1$ (left), $0.5$ (middle), $1$ (right). (Bottom row) Line cuts at $k = 0$ (left subpanels) and $k = \pi$ (right subpanels), at three temperatures $T/\Omega = 0.1$, $0.4$, $1.0$ from top to bottom within each panel. The right column ($\lambda = 1$) includes a comparison to finite-$T$ DMRG benchmark data (dashed): at $T/\Omega = 0.1$ and $0.4$, from Ref.~\onlinecite{jansen2020finite}; at $T/\Omega = 1.0$, from our own DMRG calculations, with computational details given in Appendix~\ref{app:dmrg}. Agreement is semi-quantitative across all three temperatures. Other parameters: $\Omega = 1$, $\eta = 0.05$, cloud configuration $M N$-$M_t N_t = 39$-$35$.}
\label{fig:spectral}
\end{figure*}

\begin{figure*}
\centering

\begin{minipage}[t]{0.3\linewidth}
  \centering
  \vspace{8pt}
  \makebox[1.1\linewidth][l]{\small (a)}\\
  \includegraphics[width=1.1\linewidth]{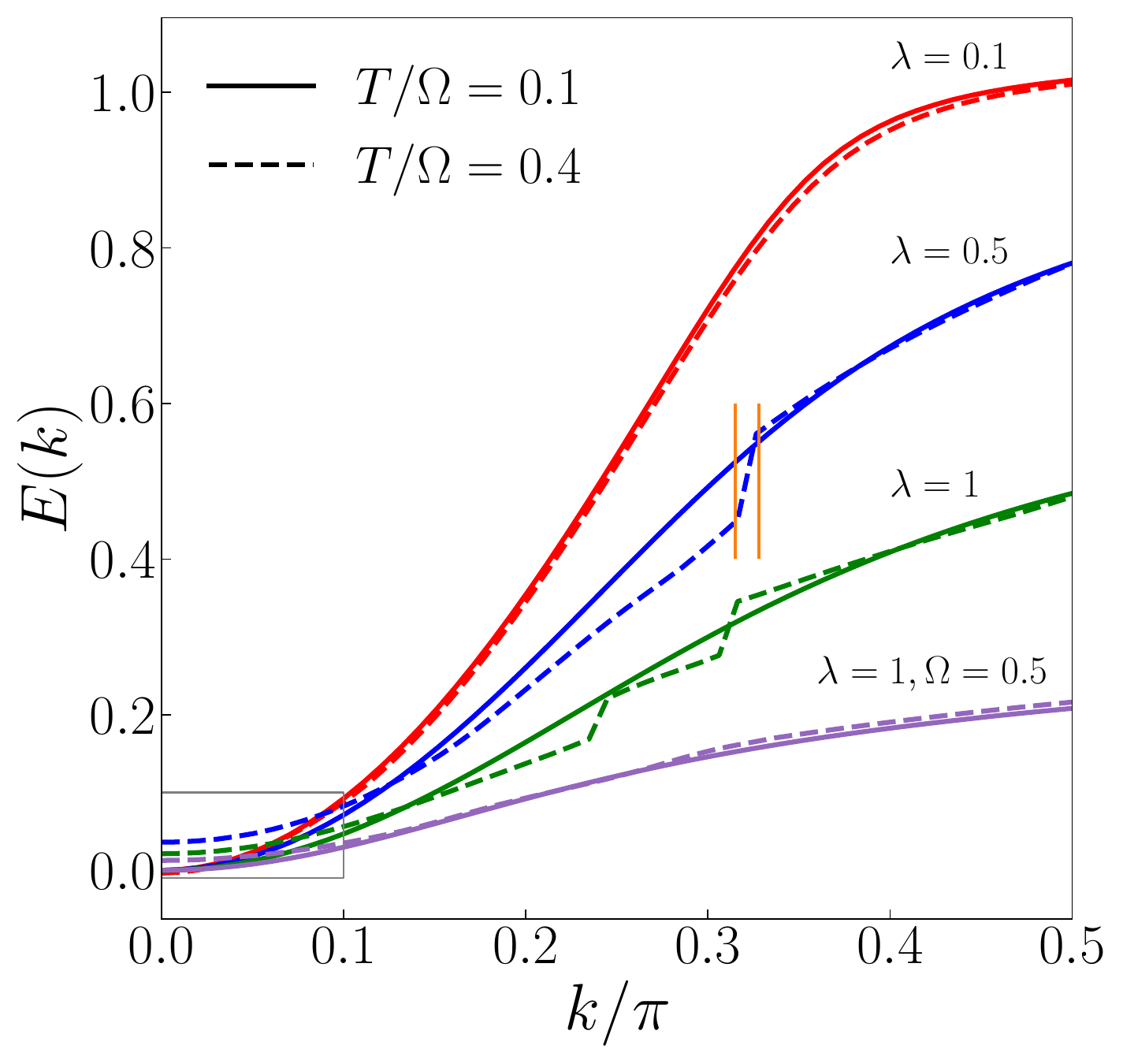}
\end{minipage}
\hfill
\begin{minipage}[t]{0.3\linewidth}
  \centering
  \vspace{8pt}
  \makebox[1.1\linewidth][l]{\small (b)}\\
  \includegraphics[width=1.1\linewidth]{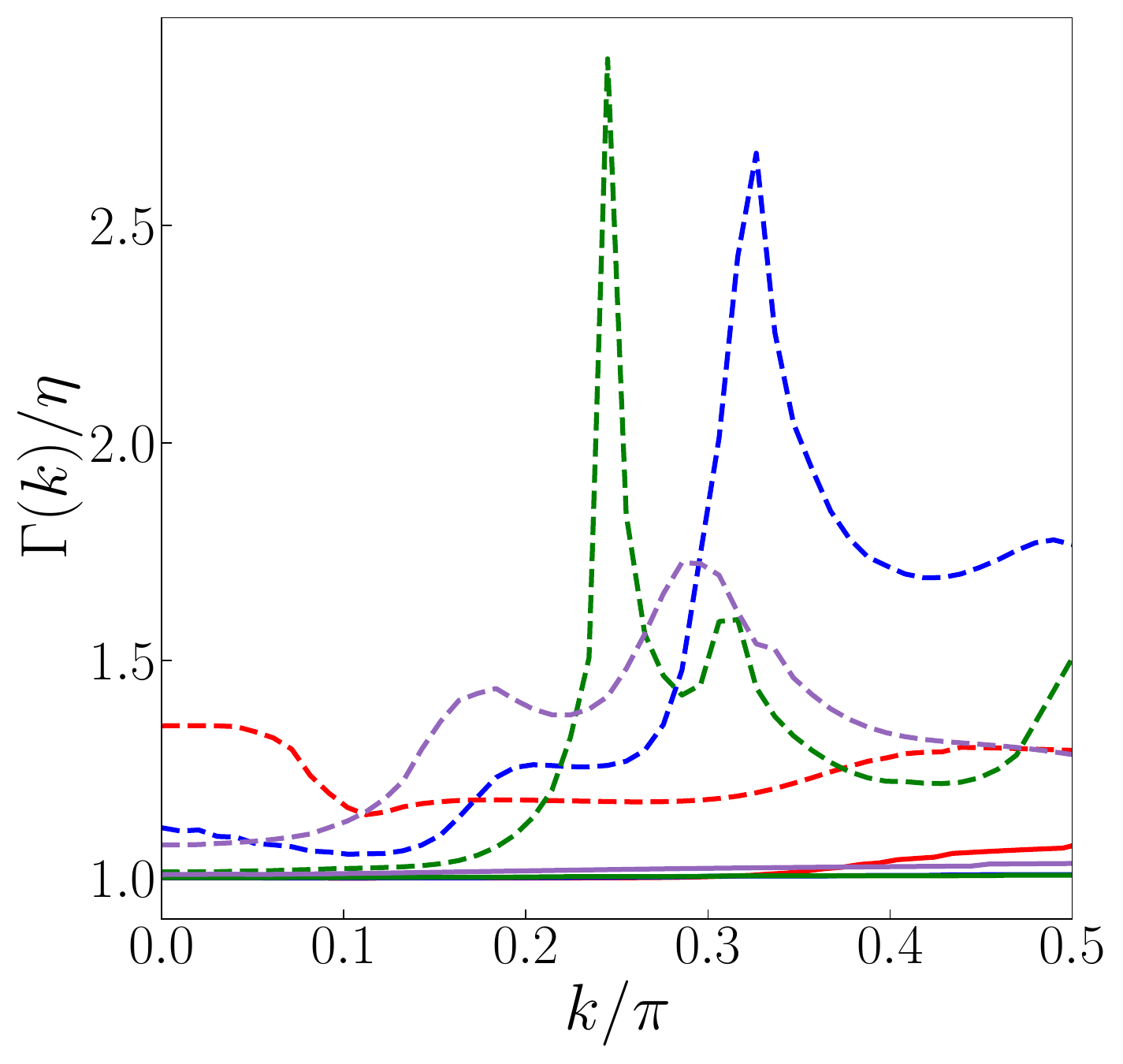}
\end{minipage}
\hfill
\hspace{-18mm}
\begin{minipage}[t]{0.37\linewidth}
  \centering
  \vspace{15pt}
  \makebox[0.56\linewidth][l]{\small (c)}\\
  \includegraphics[width=0.5\linewidth]{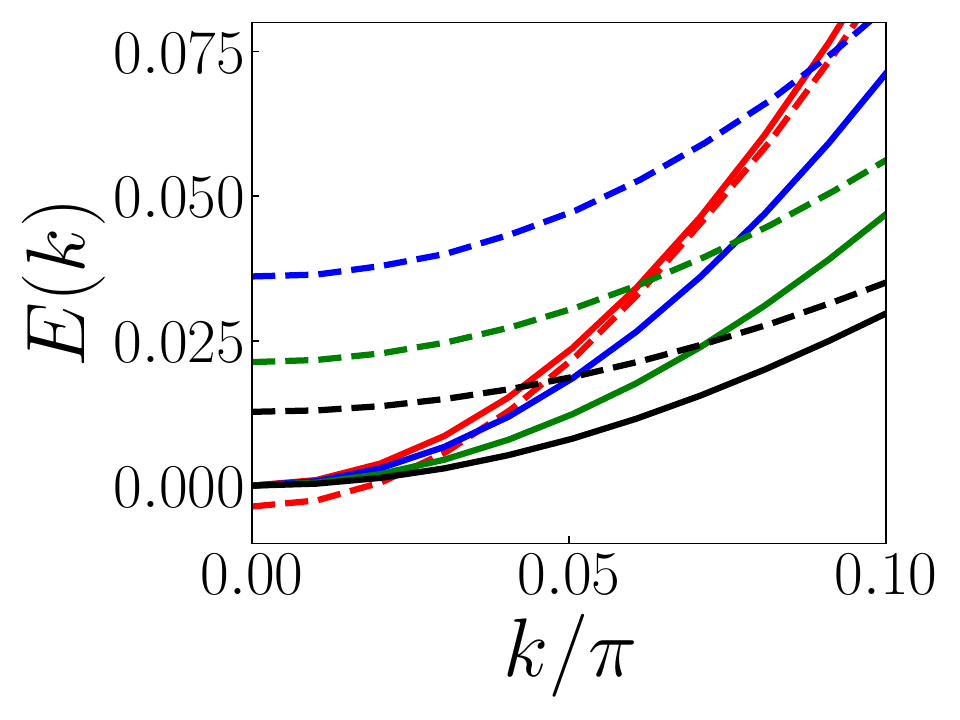}
   \vspace{0.1mm}
   
  \hspace{5mm}\makebox[0.56\linewidth][l]{\small (d)}\\
  \hspace{5mm}\includegraphics[width=0.33\linewidth]{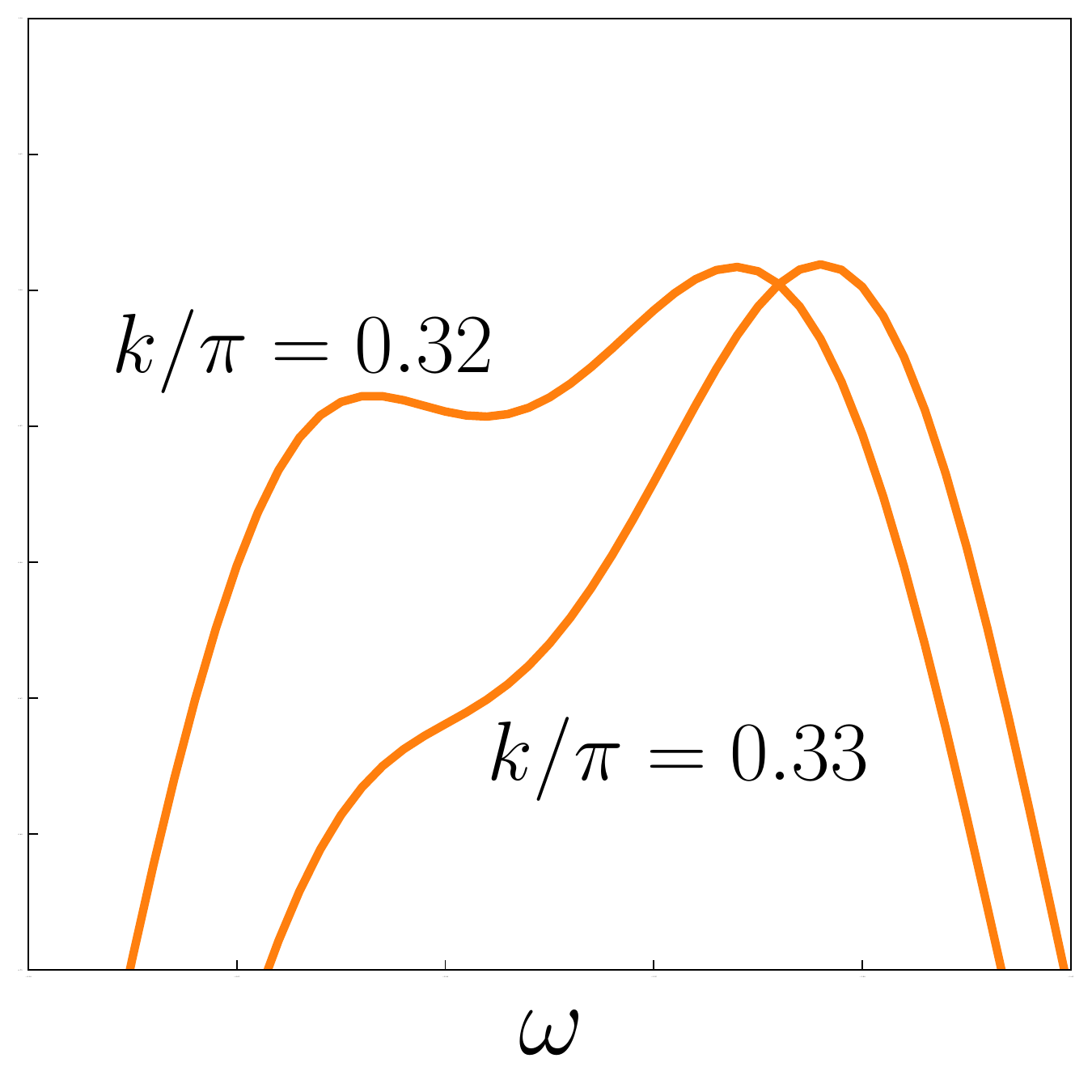}
\end{minipage}

\caption{(a) Polaron dispersions $E(k)$ for the one-dimensional Holstein model at varying $\lambda$, $\Omega$, and temperature. Phonon frequency is $\Omega = 1$ for all curves except the one explicitly labeled $\Omega = 0.5$ (purple). Solid lines: $T/\Omega = 0.1$. Dashed lines: $T/\Omega = 0.4$. (b) Polaron broadening $\Gamma(k)/\eta$, extracted from a Lorentzian fit to the polaron peak. Note that the plotted $\Gamma(k)$ is the total fitted width and includes the artificial broadening $\eta = 0.05$; the intrinsic broadening is $\Gamma(k) - \eta$ and vanishes at $T = 0$, so the solid lines saturating near unity correspond to negligible intrinsic broadening. (c) Small-$k$ zoom of panel (a), showing the band curvature near $k = 0$. (d) Line cuts of $A(k, \omega)$ at fixed $k$ across the discontinuous transition near $k/\pi \approx 0.32$ at $\lambda = 0.5$, $T/\Omega = 0.4$, marked with vertical lines in panel (a): the polaron peak develops a two-headed structure and the peak-finding procedure switches from the lower- to the upper-frequency feature.}
\label{fig:dispersion}
\end{figure*}

\section{Implementation}
\label{sec:implementation}

The TGCE method is implemented in the open-source GGCE Python package~\cite{carbone2022generalized}, building on the zero-temperature GGCE solver. The doubled Hilbert space is handled by the same multi-species cloud machinery already developed for the Holstein-Peierls model, in which the cloud-configuration vector tracks occupations of two boson species on each site. From an implementation standpoint, the only model-specific input is the squeezed interaction Eq.~\eqref{eq:V_squeezed} together with the modified non-interacting energy $\Omega(n - \tb n)$, both of which are encoded in the \code{Model} object that defines the Hamiltonian to the equation generator. The cloud parameters $(M, N, M_t, N_t)$ are exposed as user-level convergence knobs.

The equation generator returns a sparse linear system $A\, \vecx = \vecb$ whose solution yields $G(k, \omega)$ at a single $(k, \omega)$. Three solver backends are available, each with different trade-offs: a \code{scipy.sparse} direct solver, a custom continued-fraction (contfrac) solver that exploits the hierarchical block structure of the equations, and a PETSc-based sparse solver that supports massive across-points parallelism over the grid of $(k, \omega)$ points and distributed-memory matrix storage. We compare the three backends quantitatively in Sec.~\ref{sec:performance:solvers}. We have intentionally kept the discussion here high-level; further details about the implementation are documented in Ref~\cite{carbone2022generalized}.

\section{Performance and Benchmarks}
\label{sec:performance}

We characterize the computational demands of TGCE on the one-dimensional Holstein model, where benchmark finite-$T$ results from DMRG~\cite{jansen2020finite} are available for comparison. Throughout we set $t = 1$ and $\eta = 0.05$ unless otherwise specified; the dimensionless coupling is defined as $\lambda = g^2/(2t\Omega)$. We restrict the analysis to one dimension, where the free propagator admits a closed form; extension to higher dimensions is straightforward at the level of the formalism but requires numerical evaluation of $G_0$ on a momentum grid.

\subsection{Matrix scaling and sparsity}
\label{sec:performance:scaling}

We first examine the size and structure of the linear system that TGCE must solve. Figure~\ref{fig:matrscaling_size} shows the matrix dimension and the time to generate the equations of motion as a function of the cloud cutoff $N$ at fixed $N_t = N$, for several values of $(M, M_t)$. Both quantities follow power-law scaling with $N$, as is the case at zero temperature; the doubling of the boson species enters as a multiplicative prefactor. For the parameter regimes relevant in this work, equation generation is not a bottleneck. The dominant cost is in the linear solve.

The sparsity of the linear system is examined in Fig.~\ref{fig:matrscaling_sparsity}. The density $d$ (number of non-zero entries divided by the number of matrix entries) drops rapidly with increasing $N$, as expected: most cloud configurations have no direct coupling to most others. More informative for the linear solve is the edge density, $d_e \equiv \mathrm{nnz}/\sqrt{\mathrm{size}}$, which would be of order unity for a canonically sparse matrix in which the number of non-zeros grows only linearly with the dimension. We find $d_e$ to slowly \emph{grow} with cloud parameters, indicating that the number of non-zeros scales as $\mathrm{size}^{1+\gamma}$ with $\gamma > 0$, between linear and quadratic in the matrix dimension. The matrix becomes less sparse, in this sense, as TGCE is pushed to larger cloud cutoffs.

\subsection{Convergence in the cloud parameters}
\label{sec:performance:convergence}

A practical question is how rapidly TGCE converges in the four cloud parameters $(M, N, M_t, N_t)$, and whether convergence of the polaron peak is faster than convergence of the full spectrum, as is typically the case at $T = 0$. Figure~\ref{fig:convergence} addresses this. We show line cuts of $A(k, \omega)$ in a frequency window around the polaron peak at $\lambda = 1$, $\Omega = 1$, comparing several cloud configurations labeled by the notation $M N$-$M_t N_t$ (so that $39$-$35$ corresponds to $M = 3$, $N = 9$, $M_t = 3$, $N_t = 5$). At intermediate temperature $T/\Omega = 0.4$ (top row), the polaron peak is well-converged at modest cloud cutoffs at both $k = 0$ and $k = \pi$; thermal-satellite features below the polaron require slightly larger clouds but are themselves converged for $M N$-$M_t N_t = 39$-$35$ to within a few percent. At higher temperature $T/\Omega = 1$ (bottom row), the convergence is visibly slower, and the curves do not fully collapse onto a single line within the cloud sizes we examine; the polaron peak position, however, remains stable across configurations.

Some features of the convergence are unintuitive and reflect the structure of the doubled Hilbert space. At $k = \pi$ and high $T$, the convergence is more sensitive to the real-cloud size $N$ than to its fictitious counterpart $N_t$, while at $k = 0$ the dependence is the reverse. This is consistent with the picture that real and fictitious clouds enter the free propagator with opposite signs (cf.\ Eq.~\eqref{eq:Hbar_holstein}): emitting a real phonon raises the effective energy seen by $G_0$, while emitting a fictitious one lowers it. As a result the two species are not interchangeable, and the appropriate cloud cutoffs cannot be chosen independently of one another. Convergence in TGCE is therefore a four-parameter problem in a way that has no analog in zero-temperature GGCE; for the calculations that follow we have settled on $M N$-$M_t N_t = 39$-$35$ as a configuration that is well-converged at $T/\Omega \le 0.4$ and represents the best available compromise between accuracy and cost at $T/\Omega = 1$.

\subsection{Linear solver comparison}
\label{sec:performance:solvers}

Figure~\ref{fig:solver_runtime} compares the time per $(k, \omega)$ point for the three available solver backends, on a single compute node distributed across 4 MPI processes with 8 CPUs each. For small matrix sizes, the \code{scipy} backend is the fastest; for matrix sizes above $\sim 10^3$, however, its runtime grows rapidly and non-monotonically, reflecting both the cost of LU factorization on increasingly less-sparse matrices (cf.\ Sec.~\ref{sec:performance:scaling}) and the absence of fine-grained parallelism. The PETSc backend is the most performant of the three across the range of matrix sizes relevant to the calculations of this paper, outperforming continued fractions by a factor of $\sim$2--5 and the SciPy direct solver by an order of magnitude or more at the largest sizes. The PETSc backend also benefits from a basis-matrix separation feature that allows the matrix to be distributed across MPI ranks, removing a memory bottleneck that limits the other two backends.

\begin{figure}[t]
\raggedright
\hspace{4mm}\includegraphics[width=0.8\linewidth]{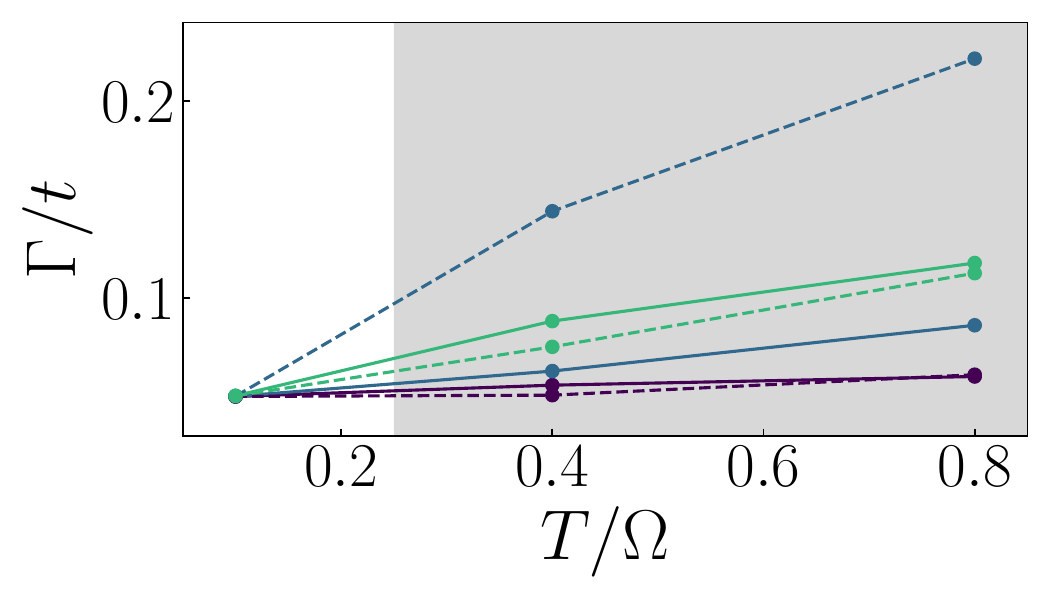}\\
\vspace{2mm}
\hspace{1.5mm}\includegraphics[width=0.84\linewidth]{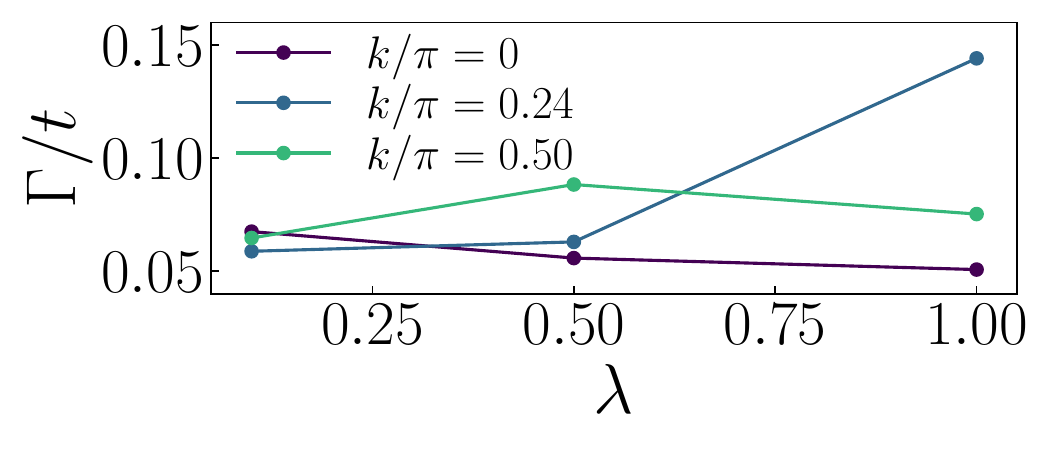}
\caption{(a) Polaron broadening $\Gamma/t$ as a function of temperature $T/\Omega$ at three wavevectors. As in Fig.~\ref{fig:dispersion}(b), the plotted $\Gamma/t$ is the fitted Lorentzian width and includes the artificial broadening $\eta/t = 0.05$; the intrinsic, physically meaningful broadening is $(\Gamma - \eta)/t$. Solid lines: $\lambda = 0.5$. Dashed lines: $\lambda = 1$. (b) $\Gamma/t$ as a function of coupling $\lambda$ at $T/\Omega = 0.4$ for the same wavevectors. Phonon frequency is $\Omega = 1$ throughout. The broadening is strongly non-uniform across the Brillouin zone, increasing rapidly with temperature near intermediate momenta where the band develops the breakdown structure shown in Fig.~\ref{fig:dispersion}, and remaining mild near $k = 0$. The shaded region in (a) marks $T/\Omega > 0.25$, where convergence is less tight.}
\label{fig:lifetime_T_lambda}
\end{figure}

\section{Results}
\label{sec:results}

We now apply TGCE to compute polaron observables for the one-dimensional Holstein model across a range of couplings $\lambda$ and temperatures $T/\Omega$. Unless otherwise noted, we use $\Omega = 1$, $\eta = 0.05$, and cloud configuration $M N$-$M_t N_t = 39$-$35$, justified by the convergence study in Sec.~\ref{sec:performance:convergence}.

\subsection{Spectral function}
\label{sec:results:spectral}

Figure~\ref{fig:spectral} summarizes the spectral function $A(k, \omega) = -\pi^{-1} \mathrm{Im}\, G(k, \omega)$ across coupling and temperature. The top row shows $A(k, \omega)$ as a function of momentum at fixed intermediate temperature $T/\Omega = 0.4$, for three coupling strengths $\lambda = 0.1$, $0.5$, and $1$. The bottom row shows line cuts at $k = 0$ and $k = \pi$ at three temperatures $T/\Omega = 0.1$, $0.4$, $1.0$ for each of the three couplings; for $\lambda = 1$, dashed curves show finite-$T$ DMRG benchmark data from Ref.~\cite{jansen2020finite} at $T/\Omega = 0.1$ and $0.4$.

At weak coupling $\lambda = 0.1$, the spectrum is close to that of the bare tight-binding band, tracing a cos-like dispersion until it merges into a continuum at $\omega \approx -2t + \Omega$. The polaron peak is sharp at all three temperatures and remains the dominant feature in the spectral window. Faint satellite peaks below the polaron grow with temperature; these are the thermal-emission satellites, in which the polaron is dressed by one or more fictitious bath phonons in the TFD picture, and which exist only at $T > 0$.

At intermediate coupling $\lambda = 0.5$, the thermal satellites are already visible at $T/\Omega = 0.4$, and a substantial fraction of the spectral weight has migrated away from the polaron band. At $T/\Omega = 1$ the polaron peak at $k = \pi$ becomes difficult to identify above the thermal background.

At strong coupling $\lambda = 1$, the spectrum at $T/\Omega = 0.4$ contains a clear replica of the polaron band offset by $+\Omega$ above the main band, with additional features at $+2\Omega$ becoming visible in the line cuts; the thermal satellites compete in weight with the polaron-band replica. The benchmark against DMRG~\cite{jansen2020finite} is favorable: at $T/\Omega = 0.1$ and $0.4$ the agreement is essentially quantitative across the spectral window, including in the thermal-satellite region. This benchmark establishes that TGCE delivers numerically accurate finite-temperature spectra in a regime where independent DMRG data are available, while returning the spectrum directly at the requested $(k, \omega)$ rather than via Fourier transform of a finite-time correlation function. We note in this context that finite-$T$ DMRG spectra are subject to their own systematic limitation: frequency resolution requires propagation to long times, and residual finite-truncation-time effects cannot be fully excluded, particularly for spectral features other than the dominant polaron peak. Residual discrepancies between the two methods at the highest temperature should be read with this in mind.

A practical comment is in order. The benefit of GGCE's per-cloud truncation, as opposed to per-site phonon-number truncation, is most pronounced precisely in the strong-coupling regime where on-site phonon occupation can become large but the cloud remains spatially compact. The same applies to TGCE: the addition of the fictitious sector enlarges the configuration space, but the cloud-truncation structure carries over, and the strong-coupling spectra in Fig.~\ref{fig:spectral} are accessed without any change of strategy from the weak-coupling case.

\subsection{Polaron dispersion and lifetime across the Brillouin zone}
\label{sec:results:dispersion}

Figure~\ref{fig:dispersion} shows the polaron dispersion $E(k)$ and the corresponding broadening $\Gamma(k)/\eta$, extracted respectively from the position and the Lorentzian width of the polaron peak in $A(k, \omega)$, across the Brillouin zone for several values of $\lambda$ and $\Omega$. A brief comment on the quantity plotted is in order. The broadening $\Gamma(k)$ extracted from the fitted Lorentzian width is the sum of the artificial broadening $\eta$ added to the propagator in Eq.~\eqref{eq:G_propagator} and the intrinsic broadening generated by the dynamics. The physical, intrinsic broadening is $\Gamma(k) - \eta$, and it is this quantity that should vanish at $T = 0$ for a stable polaron and grow with temperature as thermal phonons open up decay channels. In Fig.~\ref{fig:dispersion}(b) we plot $\Gamma(k)/\eta$ as it is fit, so the curves saturate near unity at low temperature ($T/\Omega = 0.1$, solid lines) and the rise of the dashed curves above unity is the thermally-induced intrinsic component~\footnote{Forthcoming finite-temperature momentum-average calculations verified that $\Gamma(k) - \eta$ extracted in this way converges to an $\eta$-independent quantity at small enough $\eta$}; the curves shown here are in that converged regime. Solid lines correspond to $T/\Omega = 0.1$ and dashed lines to $T/\Omega = 0.4$; for clarity, the dispersions have been shifted so that bands of the same coupling and frequency family are aligned at the lower temperature. The first inset zooms in on the small-$k$ region, where the curvature of the band controls the effective mass.

At weak coupling $\lambda = 0.1$ (red curves), the polaron band closely follows the bare cosine dispersion, with only a small flattening near $k = 0$ due to the finite-step nature of the calculation rather than a true mass enhancement; lifetimes are essentially uniform across the zone.

At intermediate coupling $\lambda = 0.5$ (blue curves), the band flattens at small momenta, signaling an increase in effective mass, and develops two discrete jumps near $k/\pi \approx 0.32$. The corresponding inset (panel (d)) shows that at this momentum the polaron peak develops a two-headed structure: any peak-finding procedure must transition from identifying the lower-frequency feature as the polaron to identifying the upper-frequency one. We emphasize that the apparent discontinuity in $E(k)$ and the sharp peak in $\Gamma(k)$ at the same momentum are therefore artifacts of imposing a single-Lorentzian description on a spectrum that has smoothly evolved into a two-peak structure; the underlying $A(k, \omega)$ varies continuously with $k$. The physical content of this feature is the redistribution of quasiparticle weight between two nearby spectral features --- the most natural microscopic origin of the second feature being hybridization of the polaron with the thermal-satellite continuum, as discussed further in Sec.~\ref{sec:results:breakdown} --- which we interpret as a precursor of the quasiparticle's breakdown: at this momentum and temperature the notion of a sharply defined polaron is absent.

At strong coupling $\lambda = 1$ (green curves) the same scenario plays out, now with the band-flattening and apparent jumps occurring at smaller momenta. Reducing the phonon frequency further to $\Omega = 0.5$ (purple) produces a much narrower band, indicating a heavy polaron at $k = 0$.

\subsection{Temperature and coupling dependence of the polaron lifetime}
\label{sec:results:lifetime}

We isolate the temperature and coupling dependence of the polaron broadening $\Gamma$ in Fig.~\ref{fig:lifetime_T_lambda}. As discussed above, the plotted $\Gamma/t$ includes the artificial broadening $\eta/t = 0.05$; the physically meaningful intrinsic broadening is $(\Gamma - \eta)/t$, which vanishes at $T = 0$ and rises with temperature. At $T/\Omega = 0.1$ in Fig.~\ref{fig:lifetime_T_lambda}(a), all three curves cluster near $\Gamma/t \approx 0.05$, consistent with a near-zero intrinsic broadening at low temperature; the spread of values at higher $T$ and across $\lambda$ in panel (b) is the thermal contribution. The top panel shows $\Gamma/t$ as a function of $T/\Omega$ for three representative momenta $k/\pi = 0$, $0.24$, $0.5$ at $\lambda = 0.5$ (solid) and $\lambda = 1$ (dashed). Across the board, temperature broadens the polaron peak, but at strikingly different rates at different momenta: at $k = 0$ the broadening is mild, while at intermediate $k$ near the dispersion's inflection point it can be several times larger. The bottom panel shows $\Gamma/t$ as a function of $\lambda$ at fixed $T/\Omega = 0.4$ for the same three momenta. The coupling dependence is also non-uniform: increasing $\lambda$ from $0.5$ to $1$ sharply increases the broadening at $k/\pi = 0.24$ but slightly decreases it at $k = 0$ and $k = \pi/2$. The shaded region in the top panel marks $T/\Omega > 0.25$, the regime where, given our cloud cutoffs, the lifetime extraction is more sensitive to convergence; results in the unshaded region are quantitatively reliable, while those in the shaded region should be read as semi-quantitative.

\subsection{Effective mass}
\label{sec:results:effmass}

\begin{figure}[b]
\centering
\includegraphics[width=0.9\linewidth]{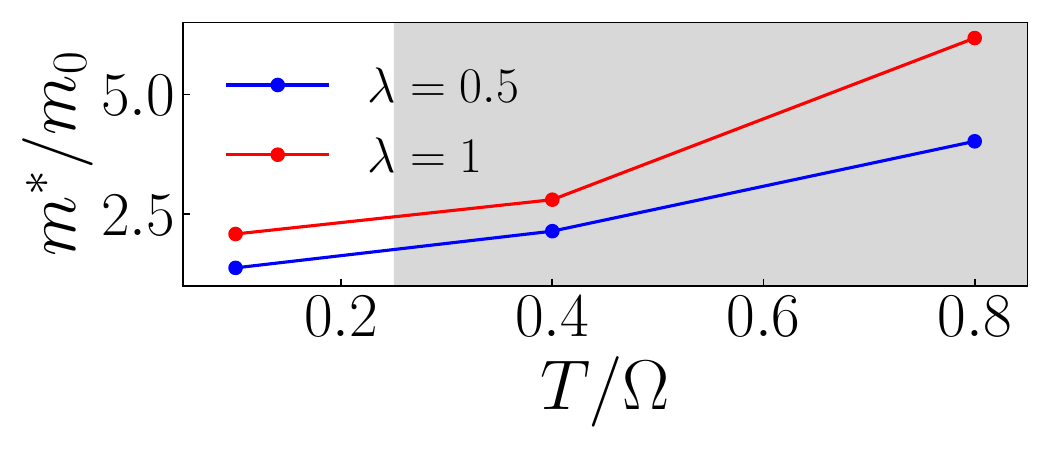}
\caption{Polaron effective mass $m^*/m_0$ near $k = 0$ as a function of temperature for $\lambda = 0.5$ and $\lambda = 1$ at $\Omega = 1$. Temperature monotonically enhances the effective mass; the shaded region $T/\Omega > 0.25$ corresponds to the regime where the lifetime and mass extraction are more sensitive to cloud-parameter convergence.}
\label{fig:effmass}
\end{figure}

Figure~\ref{fig:effmass} shows the polaron effective mass $m^* / m_0$ extracted from the band curvature near $k = 0$, as a function of $T/\Omega$, at $\lambda = 0.5$ and $1$. In both cases, increasing temperature monotonically increases the effective mass, with the rise becoming steep at $T/\Omega \gtrsim 0.4$. This effect is physically intuitive in the TFD picture: thermal dressing adds fictitious phonons to the polaron's cloud, further slowing the carrier. As above, the shaded region marks $T/\Omega > 0.25$, where the quantitative reliability of the mass extraction is reduced.

\subsection{Heavy polarons and quasiparticle breakdown at small \texorpdfstring{$\Omega$}{Omega}}
\label{sec:results:breakdown}

\begin{figure}[t]
\centering
\includegraphics[width=0.95\linewidth]{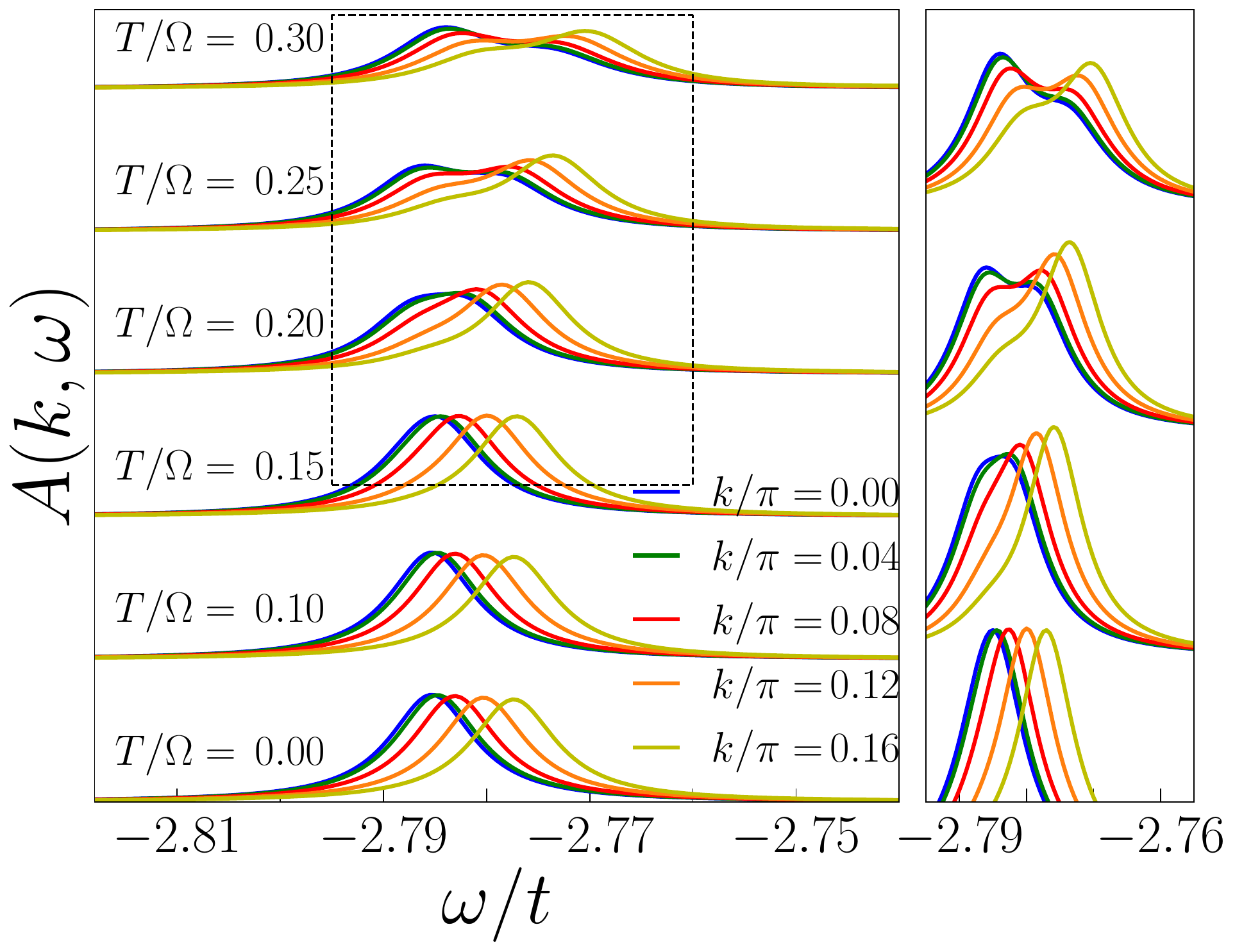}
\caption{Evolution of the polaron line shape at $\lambda = 1$, $\Omega = 0.5$ as a function of momentum (color) and temperature (vertical stacks). The polaron peak transitions from a simple Lorentzian at $T = 0$ to a multi-headed feature at the higher temperatures, signaling that the single-mode quasiparticle description ceases to be quantitatively accurate. The right panel is a zoom of the area marked by the dashed rectangle.}
\label{fig:lineshape}
\end{figure}

The most striking departure from a simple Lorentzian-broadened polaron picture occurs at small phonon frequency, where the polaron is heavy already at $T = 0$ and thermal dressing has an outsized effect on the line shape. Figure~\ref{fig:lineshape} shows $A(k, \omega)$ in a narrow window around the polaron peak at $\lambda = 1$, $\Omega = 0.5$, for several momenta $k/\pi \in [0, 0.16]$, at temperatures $T/\Omega = 0$, $0.10$, $0.15$, $0.20$, $0.25$, $0.30$. The curves are offset vertically for clarity; the right panel zooms in on the spectral window marked by the dashed rectangle in the left panel.

At $T = 0$ and the lowest temperatures, the line shape is a clean Lorentzian whose position disperses monotonically with $k$. As $T$ increases, the peak broadens; more interestingly, by $T/\Omega \approx 0.20$--$0.25$ the line shape develops a clear shoulder, and by $T/\Omega = 0.30$ the polaron peak has split into two comparable maxima. Such a structure is incompatible with a single-quasiparticle Lorentzian, and indicates that the polaron is hybridizing with nearby spectral features --- presumably the thermal-satellite continuum --- to the point that the quasiparticle weight is no longer concentrated at a single frequency. This is the regime where the very notion of a Lorentzian-broadened polaron breaks down. We emphasize that TGCE accesses this regime directly: the calculation involves no time evolution and no analytic continuation, so the multi-peak structure is a genuine feature of the computed $A(k, \omega)$ rather than an artifact of spectrum reconstruction.

\section{Conclusion and outlook}
\label{sec:conclusion}

We have introduced TGCE, a method, numerically exact in principle, for computing the momentum- and frequency-resolved single-particle Green's function of polaron problems at finite temperature, built by combining the generalized Green's function cluster expansion of Ref.~\cite{carbone2021numerically} with the thermofield double formalism~\cite{takahashi1996thermo,umezawa1982thermo}. Applied to the one-dimensional Holstein model, the method reproduces finite-$T$ DMRG benchmarks from Ref.~\cite{jansen2020finite} quantitatively across coupling regimes at temperatures up to $T/\Omega \sim 0.4$, and produces semi-quantitative spectra and observables up to $T/\Omega \sim 1$, where it is hard to converge DMRG. We have extracted polaron dispersions, lifetimes, and effective masses across coupling and temperature, identified a regime in which the simple Lorentzian-broadened polaron description begins to fail, and demonstrated the method on a parameter regime --- small $\Omega$, intermediate $T$ --- where the polaron line shape transitions to a multi-peak structure indicative of quasiparticle hybridization with thermal satellites.

Several structural features of TGCE are worth recapitulating, both as strengths and as practical limitations. On the side of strengths, the method returns $A(k, \omega)$ directly at the requested momentum and frequency. Real-time tensor-network and Lanczos-style approaches~\cite{BK,FHM,bonvca2019spectral,jansen2020finite,jansen2022finite} access the spectral function via Fourier transform of a finite-time correlation function, which restricts the spectral resolution to a window controlled by the maximum evolution time. Single-site dynamical mean-field theory~\cite{ciuchi1997dynamical,mitric2022spectral} accesses spectral functions at any frequency but loses momentum resolution unless extended to clusters at potentially significant additional cost. Diagrammatic Monte Carlo~\cite{mishchenko2015mobility,prokof1998polaron} samples in imaginary time and requires analytic continuation to obtain real-frequency spectra. TGCE is structurally complementary to all of these approaches, accessing the same physical observable through a different (and in some regimes more direct) route. The cloud-truncation strategy inherited from MA and GGCE remains well-matched to the polaron problem at finite temperature: the dominant configurations dressing the carrier remain spatially compact, even when on-site phonon occupations grow.

We note several possible limitations, which will require future improvements. First, the method is computationally expensive. The doubled Hilbert space enlarges the configuration space, and for the cloud cutoffs needed to converge the high-temperature regime in Fig.~\ref{fig:convergence}, the matrices entering the linear solve have dimensions in the $10^4$--$10^5$ range and acquire non-trivial sub-quadratic density growth with size (Fig.~\ref{fig:matrscaling_sparsity}). The PETSc backend we have used here makes the calculation tractable on commodity HPC resources, but pushing TGCE to substantially higher temperatures or to higher spatial dimensions will require either further algorithmic improvements (e.g.\ better preconditioners, low-rank approximations to the equation matrix) or substantially larger compute budgets. Second --- and we believe more importantly for users approaching the method --- the doubled phonon Hilbert space has a configuration structure that is non-trivial and, in our experience, counterintuitive. Real and fictitious bath phonons enter the free propagator with opposite-sign energy denominators (Eq.~\eqref{eq:Hbar_holstein}), so a state with $n$ real and $\tb n$ fictitious phonons produces the propagator $G_0(k, \omega - \Omega(n - \tb n))$ rather than $G_0(k, \omega - \Omega n)$. This means that the two species are not interchangeable; their cloud cutoffs must be chosen jointly, and the method is not variational in the cloud parameters, since one can lower the apparent energy of the polaron by adding fictitious phonons without bound. In practice this manifests as a four-parameter convergence problem in $(M, N, M_t, N_t)$, with non-trivial dependences between the four. We have found that the polaron peak is generally robust to these choices, but the full spectrum --- and quantities derived from the line shape, like lifetimes --- is considerably more sensitive.

Several extensions of TGCE are within reach. The generalization to higher spatial dimension requires only that the free-particle propagator be evaluated numerically; the equation-generation machinery is dimension-agnostic. Extensions to other electron-phonon vertex structures --- Peierls, mixed Holstein-Peierls, and more general short-range models --- are immediate at the formalism level and require only registering new \code{Model} objects in the implementation. Extensions to bipolarons and other multi-carrier problems, by contrast, are non-trivial: the separable initial-density-matrix assumption that underlies the TFD construction is exact for a single carrier injected into a thermalized phonon bath, but fails for two-particle response functions in non-insulating settings. We anticipate that addressing this will require either a different choice of initial state or a hybridization with diagrammatic Monte Carlo approaches that handle the inseparable-initial-state case more naturally.

Beyond methodological extensions, the central application targets of TGCE are physical settings where the finite-temperature single-particle spectrum is the observable of interest and where the relevant temperatures place the system in or near the regime $T/\Omega \sim 1$. These include charge transport in molecular and organic semiconductors, where phonon frequencies are comparable to room temperature; polaron physics in halide perovskites and other soft-lattice materials; and angle-resolved photoemission spectroscopies of correlated materials at experimental temperatures comparable to characteristic phonon scales. In all of these settings, momentum- and frequency-resolved spectra are the natural diagnostic, and a numerically exact method that returns them directly should be a useful addition to the existing toolkit.

\emph{Note added.}---During the latter stages of completion of this work, one of us (M. Berciu) and collaborators developed a complementary method for the finite-temperature Green's function of the polaron problem, based on a finite-temperature generalization of the momentum average approximation that yields a closed-form, diagrammatically derived expression for the self-energy~\cite{shannigrahi2026effective}. The main qualitative features of our results agree where they intersect. Specifically, the two works affirm that the polaron effective mass and inverse lifetime grow monotonically with temperature, and that above a temperature scale set by a fraction of the phonon frequency the quasiparticle peak can no longer be cleanly separated from the thermal background, signaling the breakdown of a simple Lorentzian quasiparticle description.

\begin{acknowledgments}
M.R.C., S.F., and J.S. contributed equally to this work. The authors thank D. Jansen, J. Bon{\v c}a, and F. Heidrich-Meisner for providing the finite-temperature DMRG benchmark data at $T/\Omega = 0.1$ and $0.4$ used in Fig.~\ref{fig:spectral}. The Flatiron Institute is a division of the Simons Foundation. Work at UBC was supported by NSERC, the Stewart Blusson Quantum Matter Institute, and the Canada First Research Excellence Fund. M.R.C. acknowledges support from the U.S. Department of Energy. We acknowledge the use of Claude (Anthropic; Claude Opus 4.7 and Claude Fable 5) for assistance with the drafting and polishing of the text in this manuscript.
\end{acknowledgments}

\emph{Disclaimer.}---This report was prepared as an account of work sponsored by an agency of the United States Government. Neither the United States Government nor any agency thereof, nor any of their employees, makes any warranty, express or implied, or assumes any legal liability or responsibility for the accuracy, completeness, or usefulness of any information, apparatus, product, or process disclosed, or represents that its use would not infringe privately owned rights. Reference herein to any specific commercial product, process, or service by trade name, trademark, manufacturer, or otherwise does not necessarily constitute or imply its endorsement, recommendation, or favoring by the United States Government or any agency thereof. The views and opinions of authors expressed herein do not necessarily state or reflect those of the United States Government or any agency thereof.

\appendix

\section{Details of the finite-temperature DMRG calculations at \texorpdfstring{$T/\Omega = 1$}{T/Omega = 1}}
\label{app:dmrg}

The finite-$T$ DMRG benchmark curves shown at $T/\Omega = 1$ and $\lambda = 1$ in Fig.~\ref{fig:spectral} were obtained from a purified matrix-product-state (MPS) simulation~\cite{verstraete2004matrix,feiguin2005finite} of the one-dimensional Holstein model, built on the same thermofield-double rotation as Eqs.~\eqref{eq:Hbar_holstein}--\eqref{eq:Vt}. The chain has $L = 21$ sites with open boundary conditions and local phonon Hilbert-space dimension $N_{\rm ph} = 21$. Real-time evolution was performed with second-order time-evolving block decimation (TEBD)~\cite{vidal2004efficient} with time step $dt = 0.05$ and singular-value truncation cutoff $10^{-7}$, with no cap on the bond dimension. A time-doubling construction~\cite{barthel2009spectral} doubles the accessible correlation window, yielding the real-space correlator on a physical-time grid $\Delta t = 0.2$ up to $t_{\rm max} = 15.6$. The correlator is transformed to the open-boundary sine modes $k_j = j\pi/(L+1)$ and Fourier transformed with the same artificial broadening $\eta = 0.05$ used in the TGCE calculations; no linear prediction is applied. The line cuts labeled $k = 0$ and $k = \pi$ in Fig.~\ref{fig:spectral} correspond to $k/\pi = 1/22$ and $21/22$, respectively. Convergence of the resulting spectra was verified against the choice of MPS algorithm and bond dimension (comparing to time-dependent variational principle calculations~\cite{haegeman2011time,haegeman2016unifying} at bond dimensions up to $\chi = 400$, using a subspace-expansion scheme~\cite{krinitsin2026subspace} closely related to the ancillary-Krylov approach of Ref.~\cite{yang2020time}, and to TEBD at truncation cutoff $10^{-8}$), the time step (down to $dt = 0.025$), the local phonon dimension (up to $N_{\rm ph} = 31$), and the system size (up to $L = 41$). Across these checks, the peak absolute residuals in $A(k, \omega)$ are at the $10^{-3}$--$10^{-2}$ level over the frequency window shown in Fig.~\ref{fig:spectral}.

\bibliography{bib}

\end{document}